\documentclass[review=false,manuscript]{acmart}

\usepackage[utf8]{inputenc}
\usepackage{csquotes}
\usepackage{hyperref}
\usepackage{todonotes}
\usepackage{booktabs}
\usepackage{siunitx}
\usepackage{textgreek}
\usepackage{threeparttable}
\usepackage{rotating}
\usepackage{multirow}
\usepackage{float}
\usepackage{tabularx}
\usepackage{array}
\usepackage{longtable}
\usepackage{pdflscape}
\usepackage{enumitem}

\begin{document}

\title{The Impact of AI Coding Assistants on Software Engineering: A Longitudinal Study}

\author{Annie Vella}
\email{annie.luxton@gmail.com}
\affiliation{%
  \institution{University of Auckland}
  \city{Auckland}
  \country{New Zealand}
}

\author{Kelly Blincoe}
\email{k.blincoe@auckland.ac.nz}
\affiliation{%
  \institution{University of Auckland}
  \city{Auckland}
  \country{New Zealand}
}

\renewcommand{\shortauthors}{Vella and Blincoe}

\setcopyright{cc}
\setcctype{by}
\acmJournal{TOSEM}
\acmYear{2026} \acmVolume{1} \acmNumber{1} \acmArticle{}
\acmMonth{1} \acmDOI{10.1145/3861343}

\begin{abstract}
AI coding assistants have become prolific in recent years. Through a longitudinal mixed-methods investigation, we examined how professional software engineers perceive the effects of AI coding assistants in regard to task focus, developer experience, and productivity. Two questionnaires were administered six months apart, yielding 158 eligible participants at the first time point, 101 at the second, and a matched longitudinal cohort of 95. 
Participants reported spending less time on most development tasks, with 82\% reporting less on writing code. We found a broader shift in focus from creation to verification activities. We propose a new category of work we term \emph{supervisory engineering work}, encompassing the direction, evaluation, and correction of AI output. This construct is broader than the extra review overhead that has been previously reported in literature since it encompasses directing and shaping AI output both before and during creation. We also identified a productivity-experience paradox: consistently positive productivity perceptions alongside eroding developer experience. 
These findings suggest that AI coding assistants may be reshaping both the nature of software engineering work and how engineers experience it.

\end{abstract}



\maketitle

\section{Introduction}
\label{chap:introduction}

AI coding assistants are changing the way software engineers work. Tools such as GitHub Copilot and Cursor, along with general-purpose AI systems like ChatGPT and Claude, have become embedded in everyday development work in just a few years~\cite{stackoverflow_ai_2024, stackoverflow_ai_2025}. 
Promising reports of productivity gains~\cite{peng_impact_2023, cui_effects_2025} have driven rapid adoption, both by organisations seeking efficiency and by individuals seeking support across a wide range of development tasks. Yet, the 2025 Stack Overflow survey found that while 60\% of developers view AI tools favourably, this represents a decline from over 70\% in previous years~\cite{stackoverflow_ai_2025}. The same survey found that more developers now distrust the accuracy of AI output than trust it, and that a majority report frustration with \enquote{AI solutions that are almost right, but not quite}~\cite{stackoverflow_ai_2025}. This mixed sentiment suggests that widespread use coexists with reservations. Beyond productivity, these tools may be reshaping where engineers direct their effort and how they experience their work.

Much of the early research in this area has centered on evaluating the technical capabilities of AI coding assistants: the correctness, security, and maintainability of generated code~\cite{yetistiren_evaluating_2023, nguyen_empirical_2022, moradidakhel_github_2023, perry_users_2023}. Other studies have examined how engineers use these tools in practice~\cite{mozannar_reading_2024, pandey_transforming_2024b, sergeyuk_using_2025}, several reporting increased review effort and a shift from producing code toward assessing it~\cite{bird_taking_2022, vaillant_developers_2024}. These accounts are snapshots, however, leaving open how the composition of engineers' work evolves with sustained use. A smaller body of work explores developers' perceptions, covering trust, usability, and impacts on learning~\cite{liang_large_2024, kuhail_will_2024}, though often with students rather than professional practitioners~\cite{vaithilingam_expectation_2022}. While productivity has been the focus of much research~\cite{houck_space_2025,weisz_examining_2024}, developer experience has received less attention despite being identified as a driver of productivity~\cite{murphyhill_what_2021,storey_towards_2021,meyer_software_2014}, and it is unclear whether that relationship holds when AI is embedded in development workflows. Further, what remains largely missing is longitudinal evidence of how professional software engineers experience these tools, and of how those experiences change.

We address this gap through a longitudinal mixed-methods study of professional software engineers using AI coding assistants. Participants completed two questionnaires six months apart, with 158 eligible respondents initially, 101 at follow-up, and 95 completing both phases. By prioritising lived experience over tool performance, this study offers a practice-oriented view of how AI coding assistants are influencing software engineering work. We examine perceptions of how focus shifts across tasks and perceptions of productivity and developer experience, guided by two research questions:

\begin{itemize}
    \item \textbf{RQ1:} How does the introduction of AI coding assistants shift software engineers' perceived focus across various development tasks, such as designing, writing, refactoring, testing, debugging, and reviewing code?
    \item \textbf{RQ2:} How does the use of AI coding assistants impact software engineers' perceptions of their developer experience and productivity?
\end{itemize}

The main contributions of this study are:

\begin{itemize}
    \item \textbf{(RQ1)} Evidence of a \emph{creation-to-verification shift}: most participants perceived spending less time on the development tasks we measured, with writing code showing the steepest decline. The balance tilted modestly toward verification activities over the study period.
    \item \textbf{(RQ1)} A proposed new category of development work, \emph{supervisory engineering work}, encompassing the effort required to direct AI, evaluate its output, and correct its errors. This construct, which emerged from convergent quantitative and qualitative patterns and encompasses more than just additional reviewing overhead, may help explain where perceived time savings from traditional tasks are being reallocated.
    \item \textbf{(RQ2)} Evidence of a \emph{productivity-experience paradox}: consistently positive productivity perceptions alongside eroding developer experience, especially in flow state and cognitive load. This pattern suggests the established relationship between developer experience and productivity may operate differently when AI is embedded in workflows, and how we define and measure developer experience may need to evolve.
\end{itemize}

Together, these findings offer a grounded perspective on how AI coding assistants may be transforming professional practice, and what that could signal for the future of software engineering.

\section{Literature Review}
\label{chap:literature-review}

This section examines existing research on the impact of AI coding assistants, spanning controlled experiments measuring productivity gains, surveys capturing developer perceptions, and studies examining workflow changes.

\subsection{Productivity and Workflow}
\label{sec:productivity-workflow}

Developer productivity has been a primary focus of AI coding assistant research. A number of controlled experiments found that task completion time reduced significantly with the use of AI~\cite{peng_impact_2023,weber_significant_2024}, as did an A/B test within a large bank~\cite{chatterjee_impact_2024} and a task-level evaluation on proprietary codebases~\cite{pandey_transforming_2024b}. Randomised field experiments running over several months measured completed tasks rather than task completion time, finding increases overall, with the largest gains among less experienced developers~\cite{cui_effects_2025}. Perceptions on productivity remain broadly positive in industry surveys~\cite{stackoverflow_ai_2024,dora_ai_2025,houck_space_2025}. Though, a study at IBM found that while productivity increased for most, some developers saw no gains at all, suggesting gains may not be uniform~\cite{weisz_examining_2024}. Another study employing a randomised controlled trial found that AI tool usage had actually \emph{increased} task completion time by 19\% even though developers estimated AI had reduced their completion time by 20\%~\cite{becker_measuring_2025}. This finding is notable but difficult to generalise due to the small sample and heterogeneous tasks.

Research has found that developers do not always complete tasks faster, due to more time spent crafting prompts and understanding, editing, and debugging AI output~\cite{vaithilingam_expectation_2022,mozannar_reading_2024}. Contributing to this, developers report difficulty steering the tools toward the desired output and that outputs are not always correct~\cite{depalma_exploring_2024, liang_large_2024}. Studies report increased review effort and describe a role shift from direct production toward assessing and integrating suggestions~\cite{bird_taking_2022,vaillant_developers_2024}. Barke et al.'s grounded theory of Copilot use distinguishes acceleration, where the developer knows what they want next, from exploration, where they are unsure and validation is more extensive~\cite{barke_grounded_2023}. Banh et al. describe an underestimated overhead in organisational adoption, arising from correcting hallucinated output, the complexity of prompting, and the cost of deployment, and report that productivity gains may be diminished by the need for manual correction~\cite{banh_copiloting_2025}. At a broader level of task allocation, a study found that access to GitHub Copilot shifted effort toward core coding activities and away from peripheral project management tasks, with effects strongest among less experienced developers~\cite{hoffmann_generative_2025}. These studies suggest that effort reallocation is occurring.

Research sheds some light on the reasons for these shifts. First, what developers delegate depends on what they enjoy: writing tests and documentation were rated least enjoyable and most likely to be delegated to AI, while implementing new features was most enjoyable and least likely to be delegated~\cite{sergeyuk_using_2025}. This aligns with broader findings that developers want AI to reduce toil, freeing them to focus on complex problem-solving~\cite{dangelo_what_2024}. Creativity in software engineering, often expressed through refactoring, designing reusable code, and learning, is something developers want to preserve~\cite{inman_seamful_2025}. Second, capability varies with task complexity and context: benchmarks show progress on isolated problems, but performance drops on multi-file, real-world issues and varies by language and difficulty~\cite{yetistiren_evaluating_2023,jimenez_swe_2023,kuhail_will_2024,nguyen_empirical_2022}.

These studies indicate that AI coding assistants redistribute engineers' focus and effort, though questions remain open. Controlled studies capture behaviour on predefined tasks, but how these patterns manifest in the varied contexts of day-to-day professional work is less well understood. Furthermore, both the underlying AI models and the tools built on them are evolving rapidly as adoption continues to grow. Understanding how engineers' focus shifts as these tools mature, and how engineers perceive those shifts, would benefit from research that tracks practitioners over time.

\subsection{Developer Experience}
\label{sec:developer-experience}

The Developer Experience (DevEx) framework identifies three dimensions: feedback loops (the speed at which developers receive information about their work), cognitive load (the mental effort required to complete tasks), and flow state (the state of full immersion and enjoyment)~\cite{noda_dev_2023}. Prior research has shown that developer experience improves productivity, satisfaction, and retention~\cite{murphyhill_what_2021,storey_towards_2021,meyer_software_2014}. This evidence base predates the widespread professional use of AI coding assistants. Whether the three dimensions remain the right ones, and whether the relationship they have with productivity still holds, is an open question in AI-assisted contexts. Studies report higher satisfaction, engagement, and enjoyment when coding with AI assistance, yet concerns exist about reduced autonomy, decreased motivation, or persistent uncertainty about correctness~\cite{chen_impact_2024,butler_dear_2025}. These trade-offs suggest that productivity gains may not always translate into improved experience.

While some research has investigated each of the DevEx dimensions in relation to AI coding assistants, the results are not conclusive. AI coding assistants may shorten \textit{feedback loops} by providing immediate suggestions, reducing pull request review cycle time, and reducing reliance on colleagues~\cite{kumar_intuition_2025,houck_space_2025}. However, the usefulness of this feedback depends on its quality and relevance, and developers still report needing to validate AI output carefully~\cite{depalma_exploring_2024}. 

Some studies found that \textit{cognitive load} was reduced from having information centralised within the IDE and eliminating context-switching to external sources~\cite{pinto_developer_2024}. However, evaluating AI outputs can also increase cognitive demands, particularly for complex tasks or when suggestions are overly verbose and lack project-specific context~\cite{alami_human_2025}. Another study found that the opacity and unpredictability of AI-generated code led to elevated stress markers and frequent interpretive confusion among novice users~\cite{zhang_neurophysiological_2025}. These findings suggest that cognitive demands may not simply decrease with AI assistance; they shift and redistribute.
How AI coding assistants affect \textit{flow state} is not yet well understood. We found only one interview study that examined this; the findings suggest that whether AI disrupts or supports flow often depends on the task at hand rather than the tool alone~\cite{lange_exploring_2025}. 

Overall, prior work indicates that AI coding assistants can improve aspects of developer experience while simultaneously introducing new sources of cognitive effort and fragmentation.

\subsection{Gaps in Current Understanding}
\label{sec:gaps}

The most significant gap is temporal. Much of the evidence described above relates to a single point in time: a survey covering five development activities~\cite{sergeyuk_using_2025}, single-session observation~\cite{barke_grounded_2023}, and one interview per participant~\cite{banh_copiloting_2025}, a limitation the last two acknowledge themselves. Controlled experiments are similarly restricted, measuring task performance within one session~\cite{peng_impact_2023,vaithilingam_expectation_2022}.

Some evidence does extend over time, but not far enough, or not in the same people. A six-week trial at a large bank surveyed participants repeatedly as they adopted an assistant~\cite{chatterjee_impact_2024}, though a structured evaluation of that length focuses on early adoption rather than sustained use. A three-week randomised trial compared developers' beliefs before and after they gained access to an assistant, and its authors recommend that future work examine longer durations, of six months to a year, to assess whether further changes emerge~\cite{butler_dear_2025}. The Stack Overflow developer survey runs annually and reported sentiment shifting considerably between 2024 and 2025~\cite{stackoverflow_ai_2024,stackoverflow_ai_2025}, which suggests perceptions do evolve with familiarity, but since each year samples a different population, it does not help us understand how an individual developer's views change over time. A systematic review of 39 studies on LLM assistants and developer productivity also suggests that most research in this space lacks longitudinal designs~\cite{mohamed_impact_2026}.

A second gap relates to population. Evidence on cognitive load under AI assistance comes from novice programming students~\cite{huang_impact_2025}, as does the closest study of usage and perceptions during a software engineering project~\cite{rasnayaka_empirical_2024}. Studies of novice programmers often rely on educational populations~\cite{prather_widening_2024}. Professional software engineering generally involves different pressures: legacy codebases, ambiguous requirements, team coordination, and real consequences for failure. Whether findings from educational contexts can be applied to professional settings is questionable.

A third gap concerns what is measured. Many studies measure what AI tools achieve rather than how engineers experience working with them, including those that track developers over time. Randomised field experiments have followed developers for several months while measuring completed tasks~\cite{cui_effects_2025}; analyses of repository activity collect no self-reports from the developers they study~\cite{xu_ai_2026,hoffmann_generative_2025}; and studies combining long-run activity data with a developer survey administer that survey once~\cite{stray_developer_2026,sergeyuk_evolving_2026}. Interview and observational studies have characterised how developers interact with these tools~\cite{barke_grounded_2023} and how generative AI is adopted in practice~\cite{banh_copiloting_2025}, rather than how the balance of activities in the work itself changes. Within experiential research, feedback loops and flow state have received much less direct attention than cognitive load. Where they appear, feedback loops are inferred from process metrics such as review cycle time~\cite{kumar_intuition_2025}, and flow state captured as one item within a broader productivity score~\cite{ziegler_productivity_2022b}, rather than studied as experiential dimensions in their own right. Productivity and developer experience are also rarely measured together in the context of AI, yet the two likely interact: an engineer who perceives productivity gains but experiences cognitive strain may develop different attitudes than one who experiences both benefits. An enterprise study of one coding assistant comes closest, pairing a large-scale survey with usability testing to examine perceptions of productivity alongside the user experience of the assistant~\cite{weisz_examining_2024}, but it captures a single snapshot and does not examine feedback loops, cognitive load, or flow state. 

These gaps motivate the research questions that guide this study. By tracking a cohort of professional software engineers over six months, we address the temporal gap directly. By focusing on practitioners rather than students, we address the population gap. By examining perceptions across task focus (RQ1) and developer experience and productivity (RQ2), we address the gap in what is measured.

\section{Methodology}
\label{chap:methodology}

This section describes the design, data collection, and analysis approach used to investigate how professional software engineers perceive the impact of AI coding assistants. 



\subsection{Study Design}
We adopted a pragmatic interpretivist stance for this research due to the socio-technical nature of the topic and the student researcher's position as a practising software engineer.

\subsubsection{Longitudinal Design}
\label{sec:study-design}

We invited the same participants to complete two questionnaires six months apart. AI coding assistants are evolving rapidly and adoption continues to grow; a single time point would only have captured a snapshot, whereas two time points allowed us to examine how perceptions evolved over time. The first questionnaire (Q1) was administered in October 2024 and the second (Q2) in April 2025, each open for four weeks.

\subsubsection{Data Collection Approach}
\label{subsec:data-collection}

Both questionnaires were administered online through Qualtrics, enabling us to reach professional software engineers globally. Each was designed to take approximately 15 minutes to complete, and followed a convergent parallel mixed-methods design~\cite{creswell_designing_2018}, with qualitative accounts contextualising quantitative patterns. 

\subsubsection{Questionnaire Structure}
\label{subsec:questionnaire-structure}

Each questionnaire had seven sections, of which four are aligned to our research questions in this paper: qualifying questions; experience and perceptions of AI tools; perceived productivity and developer experience; participant demographics. The survey also included questions on perceived impacts on skill development and self-efficacy, prompt engineering, and broader reflections on the profession; this data is not reported in this publication.
The questionnaires included Likert-scale, multiple choice, ranking, and open-ended questions to support a mixed-methods analysis. The full study instrument is provided in our online replication package~\cite{replication}.

The \emph{Qualifying Questions} section at the beginning of both questionnaires was used to check our eligibility criteria. Participants were required to be professional software engineers and to be currently using AI coding assistants. Those who answered \enquote{No} to either qualifying question were routed to the end of the questionnaire and did not complete the remaining sections.

The \emph{Experience and Perceptions of AI Tools} section captured tool usage patterns and attitudes, both providing study context as well as helping to answer RQ1. To provide context of AI use and perceptions, we asked participants which AI coding assistants they used, how frequently, and their initial impressions (in Q1) or anticipated disappointment if tools were removed (in Q2). Participants also ranked seven concerns about using AI coding assistants in their workflow: quality, readability, maintainability, security, dependency, learning curve, and other; we use \emph{primary concern} to refer to whichever category a participant ranked first. 

To answer \emph{RQ1} (How does the introduction of AI coding assistants shift software engineers' perceived focus across various development tasks?), the section also asked how participants perceived changes in focus across six core development tasks: designing, writing code, refactoring code, reviewing code, testing, and debugging. These tasks reflect core activities identified in empirical studies of software engineering practice~\cite{meyer_work_2017}. Participants were asked \enquote{When using an AI coding assistant, do you feel you spend more or less time on the following software engineering tasks?} and rated each task on a five-point scale from \enquote{Much less} to \enquote{Much more}, with \enquote{About the same} at the midpoint. The tasks were presented as labels without further definition, so participants interpreted them in the context of their own work. Participants were also asked to describe how AI coding assistants had changed their workflow.

The \emph{Perceived Productivity and Developer Experience} section captured data to answer \emph{RQ2} (How does the use of AI coding assistants impact software engineers' perceptions of their developer experience and productivity?). Participants rated their perceived productivity change and the impact of AI coding assistants on each of the three DevEx dimensions~\cite{noda_dev_2023}: cognitive load, feedback loops, and flow state. They were asked \enquote{How has your productivity changed since using an AI coding assistant?} and \enquote{How does using an AI coding assistant affect the following aspects of your Developer Experience?}, rating each on a five-point scale from \enquote{Much worse} to \enquote{Much better}, with \enquote{About the same} at the midpoint. Each of the three dimensions was further defined: feedback loops as \enquote{the speed and effectiveness with which you receive and act on feedback during development}, cognitive load as \enquote{the mental effort required to complete your development tasks}, and flow state as \enquote{the extent to which you can work seamlessly and maintain concentration without interruptions}. Productivity was deliberately not defined, so that participants would apply their own understanding of the term. They were also invited to describe specific instances where AI coding assistants helped or hindered their work in Q1, and any unexpected benefits or challenges they had experienced in Q2.

The \emph{Participant Demographics} section captured gender, age, years of professional experience, self-reported expertise, role, primary programming language, industry, company size, country of residence and country of company headquarters. In Q2, demographic questions were optional; participants could choose to update selected fields (e.g., job role or primary language) if their circumstances had changed since Q1. 

\subsubsection{Participant Recruitment}
\label{subsec:participant-recruitment}

We recruited participants using a combination of convenience and referral-chain sampling~\cite{baltes_sampling_2022}. The primary recruitment channels were our professional networks, including LinkedIn, X (formerly Twitter), Facebook, Discord, and several Slack communities, where invitations were shared directly or via community administrators. Additionally, 33 organisations were contacted to request internal distribution of which nine agreed to share the invitation with their engineering teams. Recruitment therefore drew from both individual networks and organisational channels, enabling participation from software engineers across a range of roles, experience levels, industries and countries.

\subsubsection{Ethics}
\label{subsec:ethics}

This study received approval from the University of Auckland Human Participants Ethics Committee (Reference number UAHPEC27902). All participants were provided with a Participant Information Sheet and completed an electronic consent form. 
A \$200 NZD gift card was offered as a prize draw incentive for those who completed both questionnaires. 

\subsection{Participants}
\label{sec:participants}

A total of 224 responses were recorded for Q1. Of these, 35 were incomplete (Progress $<$ 100\%), 16 respondents indicated they were not professional software engineers, and 15 reported not using AI coding assistants, yielding a final Q1 sample of 158 eligible participants (71\% eligibility rate).
All Q1 respondents who completed the questionnaire and confirmed they were professional software engineers were invited to participate in Q2, regardless of whether they reported using AI coding assistants at Q1 (${n = 173}$).
For Q2, 111 responses were recorded, of which 6 were incomplete and 4 reported not using AI coding assistants, yielding 101 eligible participants (91\% eligibility rate). 

\subsubsection{Longitudinal Attrition Analysis.}
Of the Q1 and Q2 eligible participants, 95 (60\%) participated in both and formed the matched longitudinal cohort. Fisher's exact tests compared those retained versus lost to follow-up across six variables: gender, experience level, role, programming language, company size, and country. No significant differences were found (all $p > 0.05$), and effect sizes were small (Cram\'er's $V \leq 0.23$), suggesting that attrition was not systematic.

\subsubsection{Participant Demographics}
\label{subsec:participant-demographics}

Table~\ref{tab:participant-demographics} summarises the participant demographics. Participants represented 28 countries, with the highest participation from New Zealand (41\%).

\begin{table}[!t]
\caption{Participant Demographics and Organisational Context}
\label{tab:participant-demographics}
\centering
\small
\begin{tabular}{llll}
\toprule
\textbf{Variable} & \textbf{Category} & \textbf{Q1 n (\%)} & \textbf{Q2 n (\%)} \\
\midrule

\multirow{5}{*}{\emph{Gender}}
& Man & 135 (85\%) & 86 (85\%) \\
& Woman & 19 (12\%) & 12 (12\%) \\
& Non-binary & 1 (1\%) & 1 (1\%) \\
& Self-described & 2 (1\%) & 1 (1\%) \\
& Not disclosed & 1 (1\%) & 1 (1\%) \\
\cmidrule(lr){1-4}

\multirow{6}{*}{\emph{Age}}
& 18--24 & 5 (3\%) & 2 (2\%) \\
& 25--34 & 55 (35\%) & 33 (33\%) \\
& 35--44 & 63 (40\%) & 45 (45\%) \\
& 45--54 & 28 (18\%) & 15 (15\%) \\
& 55+ & 5 (3\%) & 4 (4\%) \\
& Not disclosed & 2 (1\%) & 2 (2\%) \\
\cmidrule(lr){1-4}

\multirow{4}{*}{\shortstack[l]{\emph{Experience}\\\emph{Level}}}
& Junior & 21 (13\%) & 13 (13\%) \\
& Mid-Career & 77 (49\%) & 51 (50\%) \\
& Senior & 59 (37\%) & 37 (37\%) \\
& Not disclosed & 1 (1\%) & 0 (0\%) \\
\cmidrule(lr){1-4}


\multirow{7}{*}{\emph{Role}}
& Backend & 37 (23\%) & 28 (28\%) \\
& Frontend & 10 (6\%) & 7 (7\%) \\
& Fullstack & 57 (36\%) & 32 (32\%) \\
& Manager/Architect & 24 (15\%) & 16 (16\%) \\
& Mobile & 12 (8\%) & 7 (7\%) \\
& Specialised (e.g., Data) & 15 (9\%) & 9 (9\%) \\
& Other & 3 (2\%) & 2 (2\%) \\
\cmidrule(lr){1-4}


\multirow{5}{*}{\shortstack[l]{\emph{Company}\\\emph{Size}}}
& Small & 44 (28\%) & 25 (25\%) \\
& Medium & 44 (28\%) & 25 (25\%) \\
& Large & 34 (22\%) & 27 (27\%) \\
& Very Large & 33 (21\%) & 22 (22\%) \\
& Not disclosed & 3 (2\%) & 2 (2\%) \\
\cmidrule(lr){1-4}


\multirow{6}{*}{\shortstack[l]{\emph{Country of}\\\emph{Residence}}}
& New Zealand & 65 (41\%) & 48 (48\%) \\
& Netherlands & 16 (10\%) & 10 (10\%) \\
& United Kingdom & 14 (9\%) & 11 (11\%) \\
& Australia & 14 (9\%) & 6 (6\%) \\
& United States & 9 (6\%) & 3 (3\%) \\
& Other & 40 (25\%) & 23 (23\%) \\
\bottomrule
\end{tabular}
\end{table}

\subsection{Data Analysis}
\label{sec:analytical-structure}

\subsubsection{Overview}
Across both research questions, we followed the same three-step analytical structure. 

\paragraph{Step 1: Quantitative Cross-sectional Analysis.}
For each research question, we first analysed quantitative data from Q1 and Q2 separately. This allowed us to describe baseline perceptions at each time point. Since the Q1 and Q2 samples differ in size ($n = 158$ vs $n = 101$) and are not matched at this step, any difference between them reflects a between-sample comparison rather than within-participant change. 

\paragraph{Step 2: Quantitative Longitudinal Analysis.}
We then restricted analyses to participants who completed both Q1 and Q2. This enabled us to examine within-participant change over the study period. 

\paragraph{Step 3: Qualitative Analysis.}
We conducted a reflexive thematic analysis~\cite{braun_using_2006, braun_reflecting_2019} of the open-ended responses to understand how participants described and contextualised their experiences with AI coding assistants, complementing the quantitative results. 




\subsubsection{Quantitative Analysis}
\label{subsec:quantitative-analysis}

All quantitative analyses were conducted in R. The packages used are listed in our replication package~\cite{replication}. All analyses used complete-case data; sample sizes vary slightly across constructs due to occasional missing responses on individual items (missingness was below 1\% at both time points).

\paragraph{Measurement and Scoring.} Perceived productivity was measured with a single item, task focus with six single items, one for each development task, and developer experience with three single items, one for each dimension. In both multi-item cases, each item was analysed independently rather than combined into a scale. Single items were used to limit respondent burden across two time points, and because each item asks participants for a judgement about their own work rather than combining several indicators to infer something not directly observable. We acknowledge the limitation this imposes: internal-consistency reliability cannot be estimated for a single item. We did test whether the three developer experience dimensions could be treated as a composite and rejected it (Cronbach's $\alpha = 0.24$--$0.56$ across samples and time points), which supports analysing the dimensions separately. Single Likert items are ordinal, so we used non-parametric tests for all comparisons involving these measures. Perceptions were the object of study throughout our investigation, consistent with our pragmatic interpretivist stance. We write \emph{perceived productivity} because \enquote{productivity} can also be used to refer to output itself. Developer experience carries no such ambiguity, so we use it without qualification.

\paragraph{Descriptive Statistics}
For each outcome, we computed means, standard deviations, and response category distributions for each time point. Likert-scale responses were summarised as numeric scores (1--5) and grouped into meaningful categories (e.g., less/same/more time, positive/negative/neutral). For longitudinal data, we constructed Q1-to-Q2 transition matrices, summarised as movement patterns and visualised using alluvial diagrams.

\paragraph{Inferential Tests}
For longitudinal within-participant change, we used paired Wilcoxon signed-rank tests. For effect sizes, we report rank-biserial correlation ($r_{rb}$). 

\paragraph{Correlations}
We used Spearman rank correlations due to the ordinal nature of our Likert-scale measures. 

\paragraph{Multiple Comparison Corrections}
All multiple comparison corrections used the Holm-Bonferroni method. For cross-sectional analyses, correction was applied within each time point separately, as Q1 and Q2 represent independent cross-sections. For constructs measured through multiple related items or dimensions, corrections were applied across all items within that construct: the six task focus items (RQ1) and three developer experience dimensions (RQ2).
\label{sec:rq-extensions}

\subsubsection{Qualitative Analysis}
\label{sec:qualitative-analysis}

We used reflexive thematic analysis following Braun and Clarke~\cite{braun_using_2006, braun_reflecting_2019} to analyse the open-ended responses. The open-ended questions associated with each research question were analysed across both time points. 

\paragraph{Phase 1: Familiarisation with the Data}
\label{subsubsec:phase-1-familiarisation}

We read through all open-ended responses across both questionnaires multiple times, actively searching for meanings and patterns rather than passively absorbing content. 
These initial readings gave us a sense of the diversity of perspectives among professional software engineers and informed preliminary ideas that we carried into the coding phase.

\paragraph{Phase 2: Generating Initial Codes}
\label{subsubsec:phase-2-coding}

Using NVivo, we followed an inductive approach, generating codes from participant responses rather than applying a pre-existing framework. Codes were applied across all open-ended questions, as participants frequently discussed similar themes regardless of which question prompted them, making a shared codebook more practical than separate per-question codebooks.

To illustrate the coding process, consider the following response to the question about how AI coding assistants changed workflow:

\begin{quote}
\emph{\enquote{I find them helpful when working with languages and technologies that I am less familiar with, or use less frequently. They are often wrong but can help you move more quickly; A wrong-but-close answer can give you a jump to the next thing to search for and solve the problem.}} (P36, Q1)
\end{quote}

Rather than coding the response as a single unit, we identified distinct segments conveying different ideas. The opening clause about working with unfamiliar technologies was coded as \emph{Capability Expansion}; the observation that AI helps \enquote{move more quickly} was coded as \emph{Speed \& Automation}; and the framing of imperfect output as enabling forward momentum was coded as \emph{Confidence Boost}.

The codebook was iteratively refined throughout the coding process: new codes were added when existing codes did not adequately capture a segment, similar codes were merged to reduce redundancy, and code definitions were clarified to improve consistency. The full codebook is provided in the replication package~\cite{replication}.

For some open-ended questions, we applied an additional layer of cross-cutting valence codes to capture the evaluative dimension of participants' accounts. For example, the responses to the question asking whether AI coding assistants either helped or hindered work, each segment was coded with a valence reflecting the question wording: \emph{helpful} or \emph{hindrance}.

\paragraph{Phases 3--5: Generating, Reviewing, and Defining Themes}
\label{subsubsec:phases-3-5-themes}

Following the coding phase, we developed themes at the research-question level rather than for each open-ended question, examining coded responses from both questionnaires to identify themes spanning the longitudinal period while revealing where accounts differed between time points. Code frequency summaries helped orient this process, but themes were generated based on conceptual significance rather than prevalence. We reviewed candidate themes against the original responses and refined them through multiple iterative passes until they provided a coherent interpretation of participants' accounts. Some codes contributed to multiple themes depending on contextual framing, and where themes were multi-faceted, sub-themes captured distinct dimensions. Where similar experiences surfaced across multiple research questions, themes were shaped to reflect the specific lens of each question rather than treated as identical.

To illustrate theme generation, for RQ2, codes including \emph{Productivity \& Effort Impact}, \emph{Cognitive Load Impact}, and \emph{Prompt Interaction \& Refinement} frequently co-occurred in responses describing faster output alongside persistent friction; these were combined into a theme capturing accelerated throughput with uneven day-to-day experience.

Participants worked across a range of roles, industries, organisation sizes and levels of experience, and used different assistants at differing frequencies (Table~\ref{tab:participant-demographics}). Demographic data were imported into NVivo alongside the responses, so each segment was read alongside the respondent's stated context rather than as a free-standing statement. That context informed how participant accounts were interpreted. We did not, however, analyse themes by participant subgroup, as the sample could not support claims about differences between groups.

\paragraph{Trustworthiness}
\label{subsubsec:trustworthiness}

All coding was conducted by the first author, a practising software engineer, and that experience shaped interpretation. Reflexive thematic analysis treats this as something to be made explicit and examined rather than eliminated, and we used several reflexivity practices to do so. The first author kept reflexive memos throughout coding and returned periodically to how domain expertise might be influencing interpretation. We actively searched for counter-examples that did not fit developing themes. The second author, who did not code, took part in iterative discussion throughout the analysis rather than at theme review alone: the codebook was reviewed across several rounds of feedback, with attention to whether codes adequately captured the range of participant experiences, and candidate themes were discussed as they developed. Ambiguous segments were flagged during initial coding and revisited after the full dataset was coded to ensure consistent application. Trustworthiness in reflexive thematic analysis derives from this transparency and sustained reflexive engagement rather than from procedural checks such as independent coder agreement.

\section{Results}
\label{chap:results}

This section presents the findings from both questionnaires. To provide context for the analyses that follow, we begin by describing which AI coding assistants participants reported using, their usage patterns, and their attitudes and concerns about using these tools. We then address each RQ.

\subsection{Study Context}
\label{subsec:study-context}

\subsubsection{AI Coding Assistants in Use}
\label{subsubsec:ai-coding-assistants-in-use}

\begin{figure}[t]
\centering
\includegraphics[width=1\linewidth]{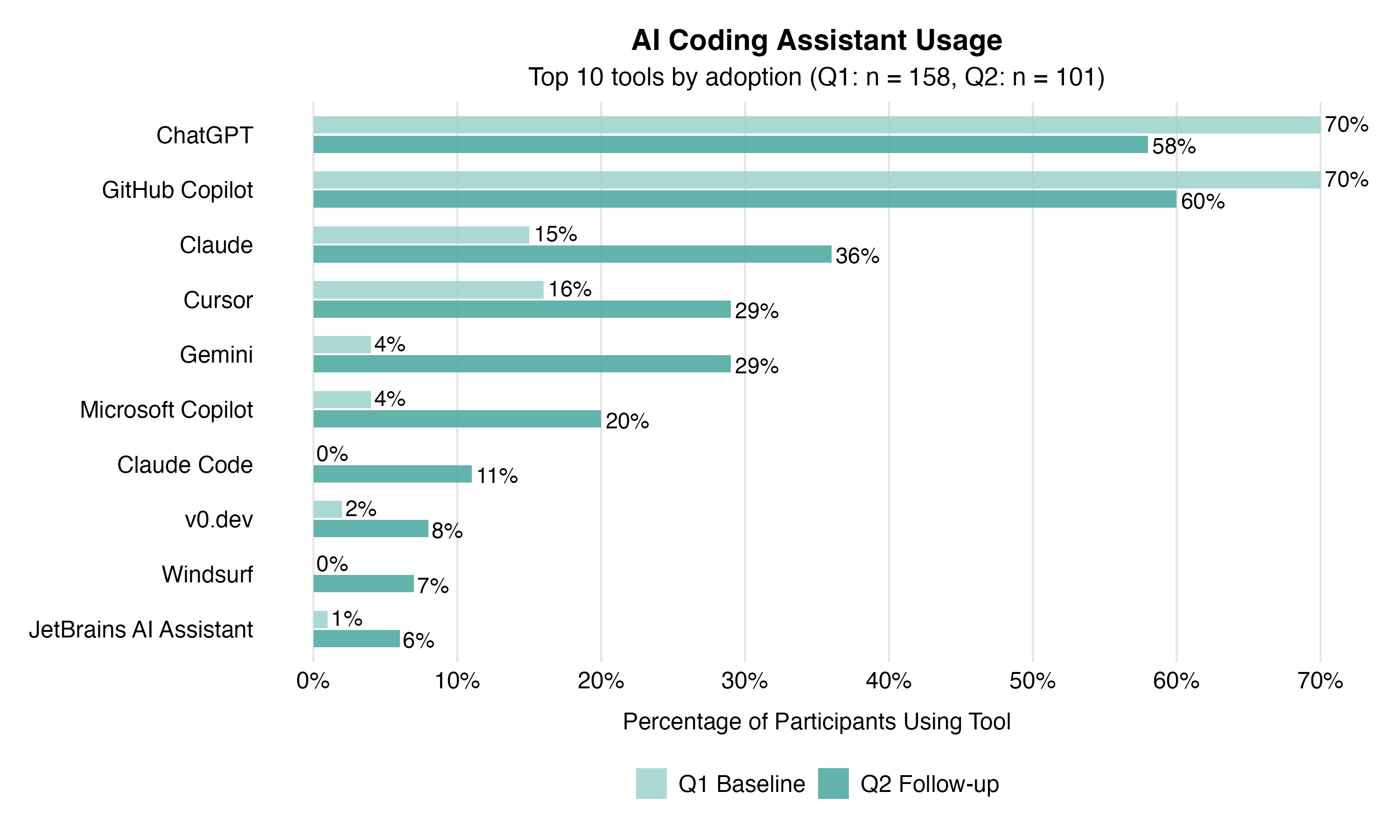}
\caption{AI coding assistant adoption rates at Q1 and Q2. Top 10 tools by maximum usage at either time point shown.}
\Description{Grouped bar chart comparing the percentage of respondents using each of the ten most-used AI coding assistants at Q1 and Q2. ChatGPT and GitHub Copilot most popular in both, but in Q2 there is a greater spread across other tools.}
\label{fig:tool-adoption}
\end{figure}

Participants reported using a wide range of AI coding assistants, with 37 unique tools mentioned across both time points (full list in replication package~\cite{replication}), including 29 at Q1 ($n=158$) and 24 at Q2 ($n=101$). GitHub Copilot and ChatGPT were the clear front-runners, both at 70\% adoption at Q1 and 58\% at Q2 (Figure~\ref{fig:tool-adoption}). However, many other tools gained in popularity between the two questionnaires.  Among matched participants ($n=95$), the tools in use evolved considerably over the study period. The number of distinct tools increased from 18 to 24, 82\% changed their tool combinations, and the mean number of tools per participant increased from 1.9 ($SD=0.8$) to 2.9 ($SD=1.6$).

\subsubsection{Usage Frequency and Location}
\label{subsubsec:usage-frequency-location}

Over half of the participants in our study used AI coding assistants daily at both time points (57\% at Q1 and 61\% at Q2). Only a small number reported using AI coding assistants once a week or less (10\% at Q1 and 8\% at Q2). Usage frequency remained stable across time points (paired Wilcoxon signed-rank, $p=0.17$, $n=95$ matched). The majority used them in both work and personal contexts (65\% at Q1, 70\% at Q2), with work-only usage less common (27\% at Q1, 24\% at Q2) and personal-only usage rare (8\% at Q1, 6\% at Q2).


\subsubsection{Attitudes Towards AI Coding Assistants}
\label{subsubsec:attitudes}

\begin{figure}[htbp]
\centering
\includegraphics[width=1\linewidth]{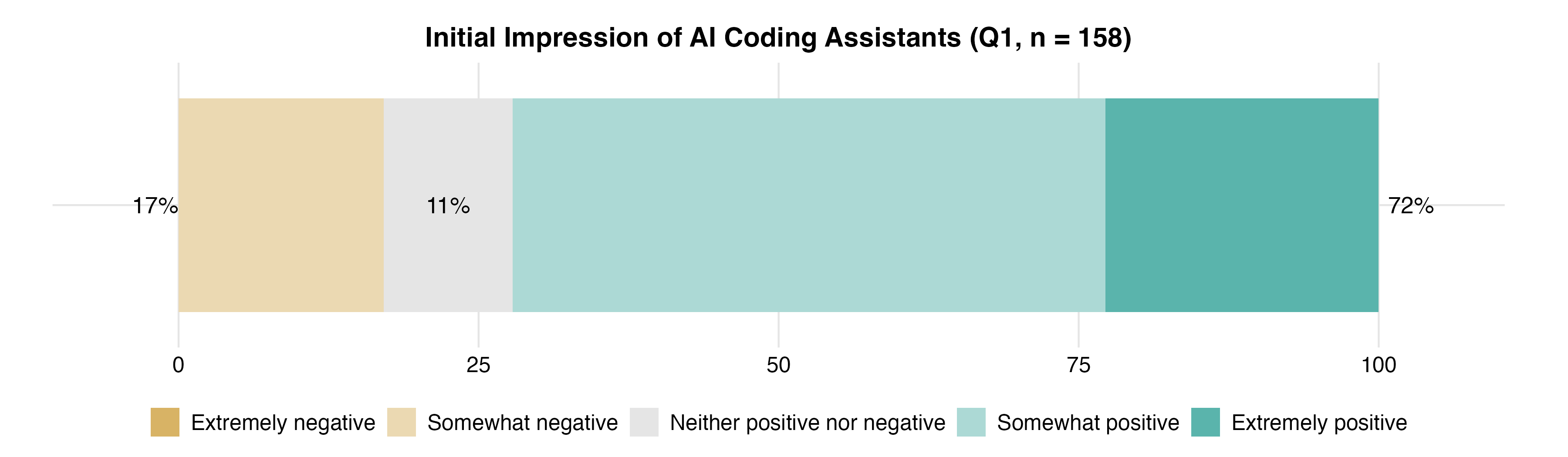}
\caption{Distribution of initial impressions of AI coding assistants at Q1.}
\Description{Chart showing initial impressions of AI coding assistants at Q1. 17\% are negative, 11\% neutral, and 72\% positive.}
\label{fig:initial-impressions}
\end{figure}

\begin{figure}[htbp]
\centering
\includegraphics[width=1\linewidth]{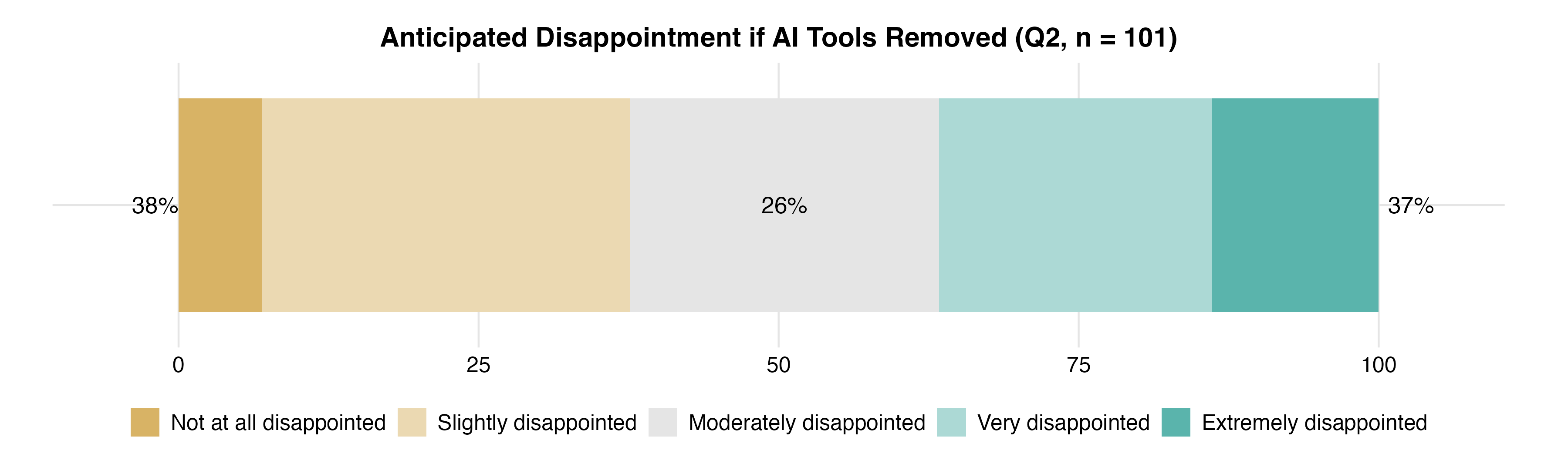}
\caption{Distribution of anticipated disappointment if AI coding assistants were removed at Q2.}
\Description{Chart showing anticipated disappointment if AI tools removed at Q2. 38\% would not be disappointed or slightly disappointed, 26\% moderately disappointed, and 37\% very or extremely disappointed.}
\label{fig:anticipated-disappointment}
\end{figure}

Participants' attitudes towards AI coding assistants were captured through two measures: initial impressions at Q1 (Figure~\ref{fig:initial-impressions}) and anticipated disappointment if tools were removed at Q2 (Figure~\ref{fig:anticipated-disappointment}). Initial impressions were predominantly positive, while anticipated disappointment was more evenly distributed. Among matched participants, the two measures showed a moderate positive correlation ($\rho=0.30$, $p=0.003$).

\subsubsection{Concerns About Using AI Coding Assistants}
\label{subsubsec:concerns}

\begin{figure}[b]
\centering
\includegraphics[width=1\linewidth]{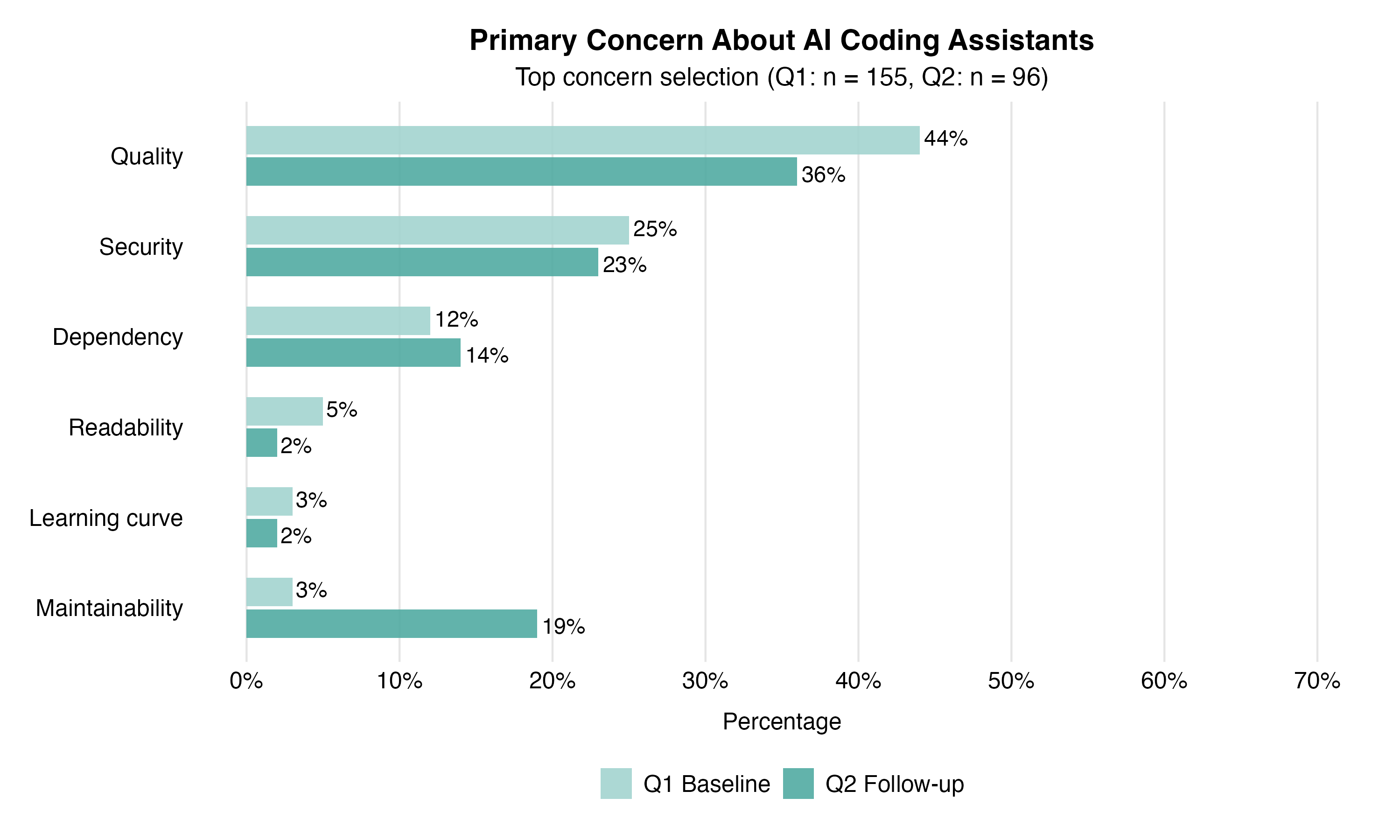}
\caption{Primary concern distribution at Q1 and Q2. \enquote{Other} excluded from analysis.}
\Description{Chart showing primary concern about AI coding assistants for Q1 and Q2. Quality most common primary concern in both, followed by security. Maintainability becomes much more of a concern in Q2 increasing from 3\% to 19\% of participants stating this as their primary concern.}
\label{fig:primary-concerns}
\end{figure}

Participants ranked seven concerns about AI coding assistants from most to least concerning. Quality was the dominant primary concern at both time points, though it declined from 44\% to 36\%, followed by security. The most notable shift was in maintainability, which rose from 3\% to 19\% as a primary concern between Q1 and Q2 (Figure~\ref{fig:primary-concerns}). Paired Wilcoxon signed-rank tests confirmed maintainability was the only statistically significant change in ranking ($n=89$ matched, $p=0.003$, moderate effect).

\subsection{RQ1: Task Focus Shifts}
\label{subsec:results-rq1}

To examine how the introduction of AI coding assistants shifts perceived task focus, we analyse perceived time reallocation across six core development activities: designing, writing, refactoring, testing, debugging, and reviewing code.
\subsubsection{Cross-sectional Analysis}
\label{subsubsec:results-rq1-baseline}

\begin{figure}[t]
\centering
\makebox[\linewidth][c]{\includegraphics[width=\linewidth]{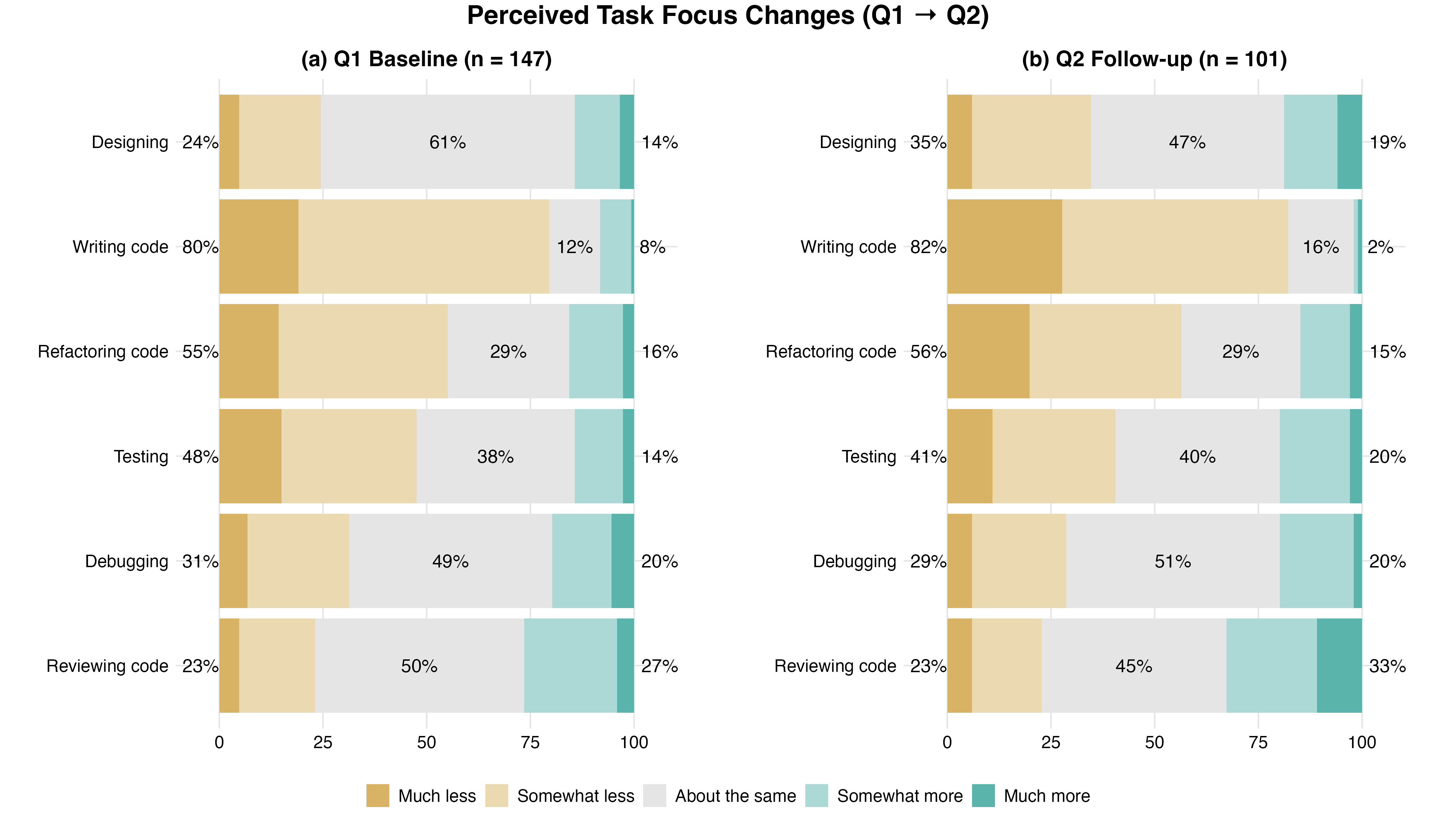}}
\caption{Task focus shifts across time points.}
\Description{Charts showing shifts in task focus at both time points. Most participants report writing less code, and refactoring less. Reviewing code was the item where the most participants reported doing more of this activity (27\% in Q1 and 33\% in Q2).}
\label{fig:task-focus-distributions}
\end{figure}

Perceived time fell across most development tasks, with writing code furthest below neutral and reviewing code the only task above it (Figure~\ref{fig:task-focus-distributions}). Writing code showed the strongest reduction at both time points ($M = 2.10$ at Q1, $M = 1.93$ at Q2), with 82\% of participants reporting less time by Q2 and only 2\% reporting more. Refactoring code and testing also sat below neutral, while designing and debugging were near neutral at both time points. Reviewing code was above the midpoint at both (Q1: 3.03, Q2: 3.15).



\subsubsection{Longitudinal Change in Task Focus}
\label{subsubsec:results-rq1-longitudinal}

Next, we examined how task focus perceptions changed within participants ($n = 88$ matched participants).
Paired Wilcoxon signed-rank tests with Holm-Bonferroni correction showed no statistically significant changes, though effect sizes varied from very small to moderate (Table~\ref{tab:longitudinal-paired-tests}).
Writing code showed the largest effect in the expected direction ($r_{rb} = -0.35$, moderate effect), while testing and reviewing showed small increases ($r_{rb} = 0.29$ and $r_{rb} = 0.26$ respectively).

\begin{table*}[t]
\centering
\caption{Longitudinal Wilcoxon Signed-Rank Tests: Changes in Task Focus Perceptions Within Matched Participants ($n = 88$)}
\label{tab:longitudinal-paired-tests}
\begin{threeparttable}
\begin{tabular}{lcccccccc}
\toprule
\textbf{Task} & \multicolumn{2}{c}{\textbf{Q1}} & \multicolumn{2}{c}{\textbf{Q2}} & & \multicolumn{3}{c}{\textbf{Test Statistics}} \\
\cmidrule(lr){2-3} \cmidrule(lr){4-5} \cmidrule(lr){7-9}
& $\boldsymbol{M}$ & $\boldsymbol{SD}$ & $\boldsymbol{M}$ & $\boldsymbol{SD}$ & $\boldsymbol{\Delta M}$ & $\boldsymbol{V}$ & $\boldsymbol{p_{adj}}$ & $\boldsymbol{r_{rb}}$ \\
\midrule
Designing & 2.94 & 0.72 & 2.89 & 0.94 & -0.06 & 595.5 & 1.000 & -0.10 \\
Writing code & 2.10 & 0.80 & 1.92 & 0.76 & -0.18 & 255.0 & 0.263 & -0.35 \\
Refactoring code & 2.48 & 0.93 & 2.39 & 1.06 & -0.09 & 688.0 & 1.000 & -0.11 \\
Reviewing code & 2.93 & 0.83 & 3.16 & 1.04 & +0.23 & 967.0 & 0.333 & 0.26 \\
Testing & 2.52 & 0.92 & 2.77 & 0.98 & +0.25 & 1030.5 & 0.263 & 0.29 \\
Debugging & 2.84 & 0.93 & 2.84 & 0.79 & 0.00 & 556.0 & 1.000 & -0.01 \\
\bottomrule
\end{tabular}
\begin{tablenotes}
\footnotesize
\item Task focus was rated on a five-point scale: 1 = much less time, 3 = about the same, 5 = much more time spent on the task when using an AI coding assistant.
\item $\Delta M$ = Mean change (Q2 - Q1). $V$ = Wilcoxon signed-rank test statistic. $r_{rb}$ = rank-biserial correlation.
\end{tablenotes}
\end{threeparttable}
\end{table*}

While aggregate statistics showed minimal change, individual transition patterns revealed substantial variation beneath these stable averages. Here we examine the three tasks with the largest effect sizes: writing code (negative) and testing and reviewing code (positive).

\begin{table*}[t]
\centering
\caption{Transition Patterns for Key Tasks ($n = 88$)}
\label{tab:task-transitions}
\begin{tabular}{lccc}
\toprule
\textbf{Task} & \textbf{Less} & \textbf{Stable} & \textbf{More} \\
\midrule
Writing code      & 30\% ($n=26$) & 56\% ($n=49$) & 15\% ($n=13$) \\
Testing           & 22\% ($n=19$) & 36\% ($n=32$) & 42\% ($n=37$) \\
Reviewing    & 23\% ($n=20$) & 38\% ($n=33$) & 40\% ($n=35$) \\
\bottomrule
\end{tabular}
\\[0.5ex]
\footnotesize Each participant's Q1 and Q2 ratings were compared. \emph{Less} and \emph{More} indicate that the rating moved down or up between time points; \emph{Stable} indicates the same rating at both. \emph{Stable} therefore means that the participant's assessment was unchanged, not that their time on the task was unchanged. Percentages may not sum to 100\% due to rounding.
\end{table*}

Table~\ref{tab:task-transitions} confirms divergent patterns: writing code showed high stability (56\%) with continued downward drift (30\% moved toward less time), while testing and reviewing showed upward movement (42\% and 40\% toward more time). Only 8\% both reduced writing time and increased reviewing time between time points. This is a constraint of the measure rather than a behavioural pattern: 81\% had already reported spending less time writing code at Q1, leaving little room for a further reduction to register.

\textbf{Creation-to-Verification Shift Analysis: }
Beyond the individual tasks, we examined whether the balance between creation- and verification-oriented work shifted. Following the Vee model~\cite{mooz_dual_2006}, we grouped the six tasks into \emph{creation tasks} (designing, writing, refactoring) and \emph{verification tasks} (reviewing, testing, debugging). These are theoretical distinctions rather than psychometric subscales, so we retain them despite low internal consistency (Creation: $\alpha = 0.40$--$0.49$; Verification: $\alpha = 0.56$--$0.57$). Table~\ref{tab:task-group-level-tests} presents the results.

Neither creation nor verification tasks changed significantly in isolation ($p_{adj} = 0.092$ and $0.060$ respectively). However, the balance between task groups showed a significant shift toward verification ($p_{adj} = 0.006$, $r_{rb} = 0.39$, moderate effect), with the mean balance score increasing from 0.26 to 0.53. When examining individual participant shifts, 42\% showed stability on at least one dimension, with the remaining 58\% shifting. Of these, decreased creation and  increased verification was the most common shift (20\%), followed by both dimensions increasing (17\%) and both decreasing (15\%).

\begin{table*}[htbp]
\centering
\caption{Task Group-Level Wilcoxon Signed-Rank Tests: Creation vs Verification Task Group Averages ($n = 88$)}
\label{tab:task-group-level-tests}
\begin{threeparttable}
\begin{tabular}{lcccccccc}
\toprule
\textbf{Task Group} & \multicolumn{2}{c}{\textbf{Q1}} & \multicolumn{2}{c}{\textbf{Q2}} & & \multicolumn{3}{c}{\textbf{Test Statistics}} \\
\cmidrule(lr){2-3} \cmidrule(lr){4-5} \cmidrule(lr){7-9}
& $\boldsymbol{M}$ & $\boldsymbol{SD}$ & $\boldsymbol{M}$ & $\boldsymbol{SD}$ & $\boldsymbol{\Delta M}$ & $\boldsymbol{V}$ & $\boldsymbol{p_{adj}}$ & $\boldsymbol{r_{rb}}$ \\
\midrule
Creation & 2.51 & 0.58 & 2.40 & 0.64 & -0.11 & 814.5 & 0.092 & -0.24 \\
Verification & 2.77 & 0.66 & 2.92 & 0.69 & +0.16 & 1527.5 & 0.060 & 0.30 \\
Balance (V-C) & 0.26 & 0.58 & 0.53 & 0.83 & +0.27 & 2365.5 & 0.006 & \textbf{0.39}** \\
\bottomrule
\end{tabular}
\begin{tablenotes}
\footnotesize
\item Each group score is the mean of its three constituent task scores.
\item $\Delta M$ = Mean change (Q2 - Q1). Balance = Verification - Creation. $V$ = Wilcoxon signed-rank test statistic. $r_{rb}$ = rank-biserial correlation.
\item Significance: *$p < 0.05$, **$p < 0.01$, ***$p < 0.001$. Holm-Bonferroni corrected.
\end{tablenotes}
\end{threeparttable}
\end{table*}

\subsubsection{Qualitative Perspectives on Workflow Changes}
\label{subsubsec:results-rq1-qualitative}

Reflexive thematic analysis of responses describing how AI coding assistants had changed participants' workflows, identified four themes capturing changes in their interactions with code, tools, and information sources.

\paragraph{Theme 1: Compressing routine creation work.}
Participants across both time points consistently described AI coding assistants as reducing the effort required for routine implementation tasks. Their accounts locate where the perceived reduction in writing code came from: boilerplate, repetitive structures and small functions.

\emph{Sub-theme 1A: Reducing manual implementation effort.}
Participants at both time points highlighted how AI tools helped them generate boilerplate, small functions, or repetitive structures. As one participant explained, \emph{\enquote{I spend less time writing repetitive and boilerplate code}} (P25, Q1) and \emph{\enquote{Mostly, the routine tasks like coding and testing now require less manual effort.}} (P98, Q2). Others noted that they now typed less: \emph{\enquote{Reduced my keystrokes. I mostly use AI only for line completion and most of the time I only have to type less than half a line before AI figures out what I want}} (P44, Q1). 

\emph{Sub-theme 1B: Supporting small-scale debugging and error handling.}
Participants described targeted improvements in resolving smaller issues, such as obtaining explanations of error messages or addressing minor test failures. This pattern appeared consistently across both questionnaires. As one participant noted, \emph{\enquote{I give the error message directly to ChatGPT and get a primer on what could be the possible cause of the issue}} (P135, Q1). Another wrote, \emph{\enquote{Troubleshooting and debugging become faster.}} (P91, Q1). 

\paragraph{Theme 2: Rerouting information seeking and comprehension into AI interactions.}
Participants described noticeable changes in how they acquired information, understood unfamiliar code, and learned new concepts. This theme appeared strongly at Q1 and remained clear at Q2, although Q2 participants often described these behaviours as normalised rather than novel. Information seeking was not one of the six rated activities, so this change sits outside the task measures.

\emph{Sub-theme 2A: From search engines to real-time assistance.}
Developers increasingly used AI as the first step when encountering errors, clarifying APIs, or checking implementation details, for example: \emph{\enquote{Instead of going to Stack Overflow for a coding problem, I now go to ChatGPT first, even before any search engine.}} (P135, Q1) and \emph{\enquote{Instead of googling the syntax for one language, I immediately ask ChatGPT to provide an example}} (P149, Q2). For many, AI effectively displaced older search habits: \emph{\enquote{It has basically replaced google for me}} (P143, Q1). By Q2 this was described as settled practice: \emph{\enquote{AI coding Assistants are my first source of information.}} (P174, Q2).

\emph{Sub-theme 2B: Accelerated learning, onboarding, and code understanding.}
Participants in both time points used AI to accelerate understanding across unfamiliar languages, frameworks, or legacy codebases. For example, one participant wrote: \emph{\enquote{Much easier to pick up new tools, libraries, languages and get productive.}} (P132, Q1). Another wrote: \emph{\enquote{It has been particularly helpful in explaining legacy code and detecting bugs more efficiently.}} (P2, Q2).  Others reported gaining conceptual clarity: \emph{\enquote{Definitely felt like it's given me a level up on understanding new concepts}} (P5, Q2).

\paragraph{Theme 3: From doing to supervising AI-generated code.}
Across both time points, engineers described shifting effort from directly producing code to supervising AI-generated output. For example, \emph{\enquote{I ask AI to make changes first and review them before accepting or rejecting them}} (P82, Q1). In the quantitative results this appears as a shift in the creation-to-verification balance rather than as an increase in reviewing code.

While some participants framed this shift in terms of \enquote{design} or \enquote{higher-level thinking}, their descriptions more consistently reflected supervision-oriented activities, such as directing the assistant, evaluating its suggestions, and deciding what to accept, modify, or discard. This supervisory work replaced portions of hands-on implementation effort but does not map cleanly onto traditional software development task categories. As one participant explained, \emph{\enquote{Sometimes, I'll not be happy with the original reference, and will 'coach' the AI to produce something that works better for my use-case}} (P75, Q1). 
At Q2, this supervisory framing became slightly more prominent. Participants increasingly articulated the effort involved in directing and controlling the assistant itself. One described their role as \emph{\enquote{Mostly reading the code and directing AI on the right way}} (P54, Q2).


\paragraph{Theme 4: Verification and trust calibration moderating perceived benefits.}
Participants varied in how much they trusted or relied upon AI-generated code. Some reported substantial workflow improvements, while others emphasised the need to verify every suggestion or restrict use to low-risk contexts. This theme was evident at both time points. One participant described this careful approach succinctly: \emph{\enquote{I almost never let it write production code, or at least never leave in code that I don't understand and or wouldn't have written myself}} (P119, Q1). Others stressed the limits of trust, particularly for complex or critical work: \emph{\enquote{I do not trust it for strategic or critical code.}} (P8, Q1).

At Q2, similar concerns remained evident, though participants reflected on how their use of AI had evolved over time. One participant noted, \emph{\enquote{I think 6 months ago, I was using the tools a lot more cautiously with a lot more fact checking}} (P109, Q2), while another highlighted the increasing verification effort required: \emph{\enquote{has increased the `is this code correct', `is it maintainable', `is it secure' and `is this just probabilistic hallucination' time up.}} (P88, Q2). The Q2 accounts moved in both directions: some participants described becoming less cautious over time, while others described a growing verification burden. 

\subsubsection{Summary}
\label{subsubsec:results-rq1-summary}

A majority of participants (69\%) perceived spending less time overall on the six development tasks at both time points. Writing code showed the largest reduction, with 82\% reporting less time by Q2 and means well below neutral at both time points ($M = 2.10$ at Q1; $M = 1.93$ at Q2). All tasks except reviewing showed means below neutral, with writing code furthest from the midpoint. Testing and reviewing were the only tasks with positive effect sizes, though neither change was statistically significant. Among matched participants, a significant shift toward verification activities emerged ($r_{rb} = 0.39$, $p_{adj} = 0.006$, moderate effect). Participants' accounts contextualise this quantitative pattern by describing compressed routine implementation, rerouted information seeking, a new supervisory layer of directing and evaluating AI output, and ongoing verification and trust calibration that moderated perceived benefits.

\subsection{RQ2: Developer Experience and Productivity}
\label{subsec:results-rq2}

To answer RQ2, we analyse two related constructs: developer experience (cognitive load, feedback loops, and flow state) and perceived productivity.

\subsubsection{Developer Experience}
\label{subsubsec:results-rq2-devex}

\begin{figure}[!tb]
    \centering
    \includegraphics[width=1\linewidth]{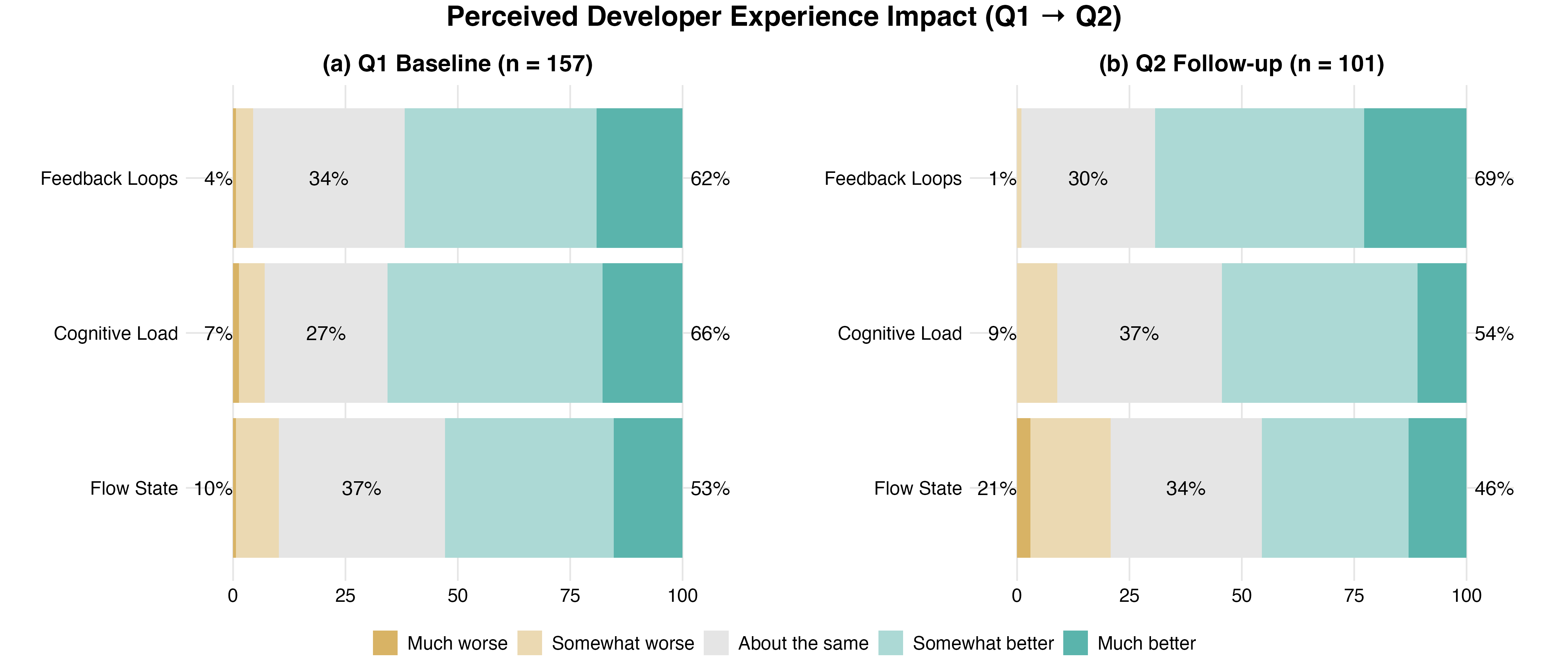}
    \caption{Perceived impact on developer experience across time points.}
    \Description{Charts showing perceived impact on developer experience across time points. At Q1, most participants report improvements for all three items (feedback loops, cognitive load, and flow state). At Q2, most still report improvements for feedback loops and cognitive load, but flow state falls to 46\%, and the share reporting a negative impact on flow state increases from 10\% to 21\%.}
    \label{fig:developer-experience-evolution}
\end{figure}

\paragraph{Cross-sectional analysis}
The three developer experience dimensions were closely grouped at Q1 but more widely separated at Q2, with feedback loops rated highest and flow state lowest (Figure~\ref{fig:developer-experience-evolution}). At Q1, participants reported similar improvements in cognitive load ($M = 3.75$) and feedback loops ($M = 3.76$), while flow state lagged behind ($M = 3.57$) and attracted more negative ratings. By Q2, feedback loops showed the greatest improvement ($M = 3.91$) while cognitive load ($M = 3.56$) and flow state ($M = 3.35$) showed smaller improvements. One-sample Wilcoxon signed-rank tests showed that all three dimensions were rated above the neutral midpoint at both time points ($p_{adj} < 0.01$), with effect sizes ranging from moderate (flow state at Q2: $r_{rb} = 0.43$) to large (feedback loops at Q2: $r_{rb} = 0.98$).


We also examined relationships between developer experience and attitudes towards AI tools, since attitudes may shape how developers use these tools and perceive their impact. Table~\ref{tab:rq2-devex-attitude-correlations} shows Spearman correlations between attitudes towards AI tools and developer experience dimensions. Both initial impressions at Q1 and anticipated disappointment at Q2 showed significant positive associations with all three dimensions, with feedback loops consistently showing the strongest association (moderate effect at both time points). Positive correlations indicate more positive attitudes are associated with better developer experience.

\begin{table}[t]
\centering
\caption{Spearman Correlations Between Attitudes and Developer Experience Dimensions}
\label{tab:rq2-devex-attitude-correlations}

\begin{tabular}{lcccc}
\toprule
\textbf{Dimension} & \multicolumn{2}{c}{\textbf{Q1 ($n=157$)}} & \multicolumn{2}{c}{\textbf{Q2 ($n=101$)}} \\
& \multicolumn{2}{c}{\textbf{Initial Impressions}} & \multicolumn{2}{c}{\textbf{Anticipated Disappointment}} \\
\cmidrule(lr){2-3} \cmidrule(lr){4-5}
 & $\boldsymbol{\rho}$ & $\boldsymbol{p_{adj}}$ & $\boldsymbol{\rho}$ & $\boldsymbol{p_{adj}}$ \\
\midrule
Feedback Loops & \textbf{0.32}*** & $<$0.001 & \textbf{0.33}** & 0.002 \\
Flow State & \textbf{0.27}** & 0.001 & \textbf{0.26}* & 0.015 \\
Cognitive Load & \textbf{0.22}** & 0.007 & \textbf{0.23}* & 0.021 \\
\bottomrule
\end{tabular}%

\begin{tablenotes}
\centering
\footnotesize
\item *$p < 0.05$, **$p < 0.01$, ***$p < 0.001$ (Holm-Bonferroni corrected)
\end{tablenotes}
\end{table}

\paragraph{Longitudinal Change in Developer Experience}

Next, we examined how developer experience perceptions changed within participants ($n = 94$ matched participants). Paired Wilcoxon signed-rank tests with Holm-Bonferroni correction revealed divergent trajectories across dimensions (Table~\ref{tab:dx-longitudinal-change}). Only feedback loops improved significantly; cognitive load and flow state showed non-significant declines.
Individual variation was substantial: for feedback loops, 33\% improved and 18\% declined; for cognitive load, 19\% improved and 29\% declined; for flow state, 27\% improved and 35\% declined.


\begin{table*}[htbp]
\centering
\caption{Longitudinal Wilcoxon Signed-Rank Tests: Changes in Developer Experience Dimensions Within Matched Participants ($n = 94$)}
\label{tab:dx-longitudinal-change}
\begin{threeparttable}
\begin{tabular}{lcccccccc}
\toprule
\textbf{Dimension} & \multicolumn{2}{c}{\textbf{Q1}} & \multicolumn{2}{c}{\textbf{Q2}} & & \multicolumn{3}{c}{\textbf{Test Statistics}} \\
\cmidrule(lr){2-3} \cmidrule(lr){4-5} \cmidrule(lr){7-9}
& $\boldsymbol{M}$ & $\boldsymbol{SD}$ & $\boldsymbol{M}$ & $\boldsymbol{SD}$ & $\boldsymbol{\Delta M}$ & $\boldsymbol{V}$ & $\boldsymbol{p_{adj}}$ & $\boldsymbol{r_{rb}}$ \\
\midrule
Feedback Loops & 3.73 & 0.78 & 3.95 & 0.74 & +0.21 & 810.5 & 0.038 & \textbf{0.38}* \\
Cognitive Load & 3.74 & 0.80 & 3.60 & 0.79 & $-$0.15 & 382.5 & 0.195 & $-$0.26 \\
Flow State & 3.54 & 0.84 & 3.36 & 1.01 & $-$0.18 & 651.5 & 0.195 & $-$0.24 \\
\bottomrule
\end{tabular}
\begin{tablenotes}
\footnotesize
\item Developer experience was rated on a five-point scale: 1 = much worse, 3 = about the same, 5 = much better, describing how using an AI coding assistant affects each dimension.
\item $\Delta M$ = Mean change (Q2 - Q1). $V$ = Wilcoxon signed-rank test statistic. $r_{rb}$ = rank-biserial correlation.
\item Significance: *$p < 0.05$, **$p < 0.01$, ***$p < 0.001$.
\end{tablenotes}
\end{threeparttable}
\end{table*}

\begin{figure}[tb]
\centering
\includegraphics[width=1\linewidth]{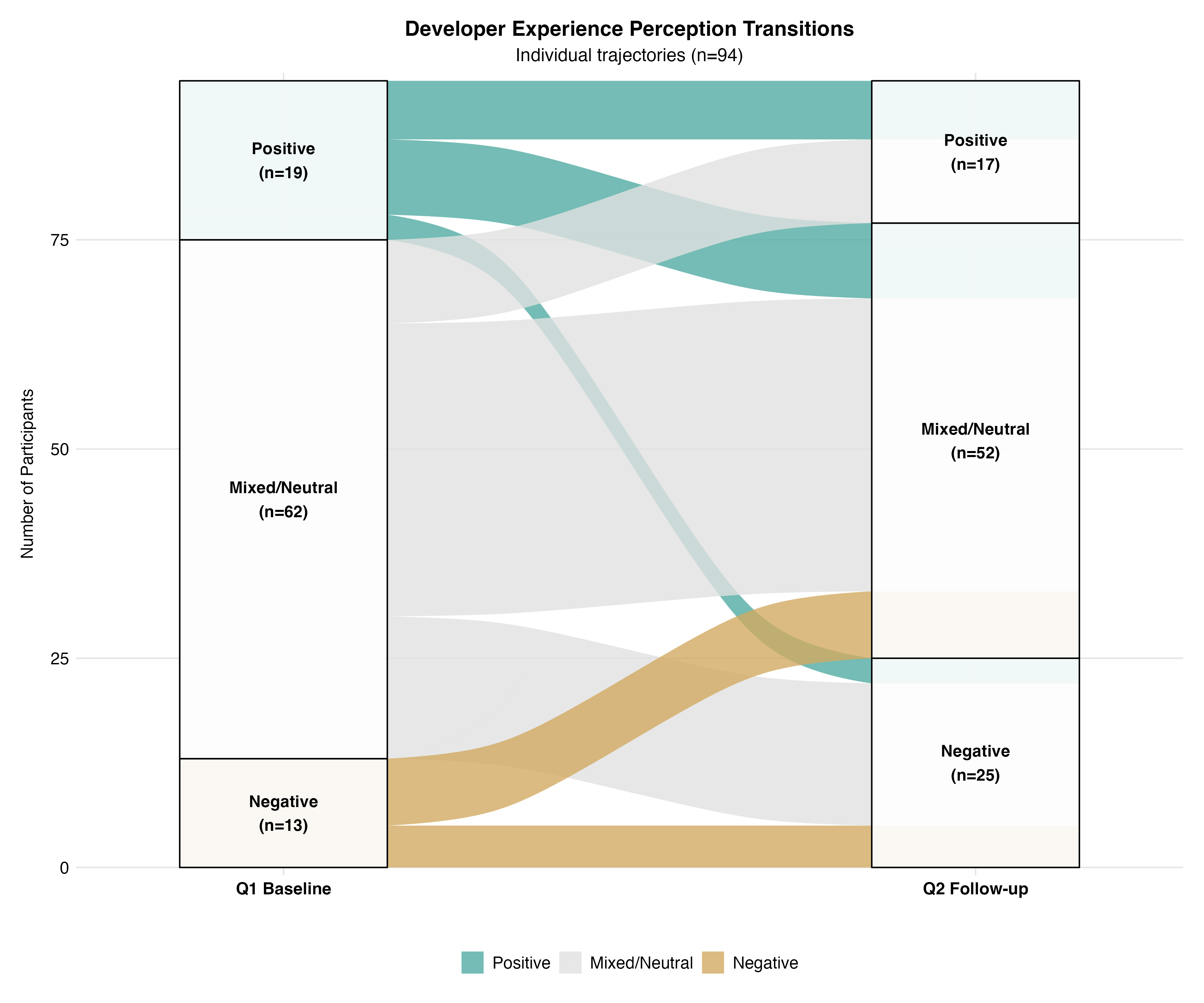}
\caption{Developer experience perception transitions between Q1 and Q2.}
\Description{Chart showing transitions between developer experience perceptions across the two time points. Of those positive at Q1, most remain positive or become mixed or neutral, and a few become negative. Of those mixed or neutral at Q1, many stay mixed or neutral, but many also shift, more often to negative than to positive. Of those negative at Q1, some stay negative and most become mixed or neutral. None move from negative at Q1 to positive in Q2.}
\label{fig:dx-cohort-transitions}
\end{figure}

Among matched participants, we classified perceptions into cohorts based on meaningful thresholds: \emph{Positive} (all three dimensions rated 4--5), \emph{Negative} (any dimension rated 1--2), and \emph{Mixed/Neutral} (neither fully positive nor containing any negative ratings). Figure~\ref{fig:dx-cohort-transitions} visualises how participants transitioned between these cohorts over the study period. The cohort analysis reveals a drift toward negative developer experience. The Negative cohort nearly doubled from 14\% to 27\%, while the Positive cohort showed poor retention at 37\%. No participant who started in the Negative cohort recovered to fully Positive by Q2. Complete stability across all three dimensions was rare (9\%). 
Within the Negative cohort, flow state was the predominant issue, affecting 54\% at Q1 and rising to 76\% at Q2.

\subsubsection{Productivity}
\label{subsubsec:results-rq2-productivity}

\paragraph{Cross-sectional analysis}

\begin{figure}
    \centering
    \includegraphics[width=1\linewidth]{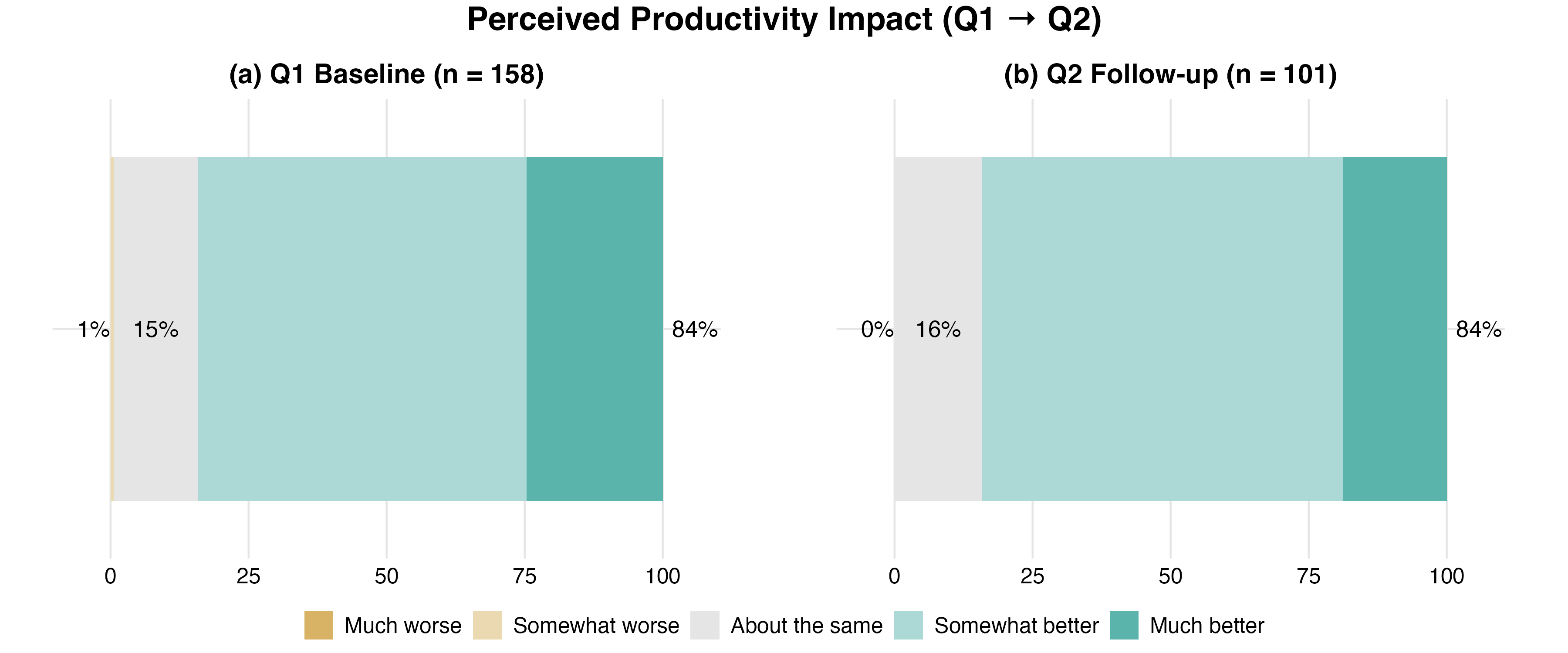}
    \caption{Perceived productivity impact across time points.}
    \Description{Chart showing perceived productivity impact at both time points. At both, most participants (84\%) report productivity is better.}
    \label{fig:productivity-evolution}
 \end{figure}

At Q1, productivity perceptions were predominantly positive ($M = 4.08$, $SD = 0.65$), with 84\% reporting improvement. At Q2, perceptions were similar ($M = 4.03$, $SD = 0.59$), with identical 84\% reporting improvement (Figure~\ref{fig:productivity-evolution}). 

Spearman correlations showed a significant moderate positive association between initial impressions and productivity at Q1 ($\rho = 0.31$, $p < 0.001$, moderate effect). At Q2, anticipated disappointment showed a stronger positive correlation ($\rho = 0.48$, $p < 0.001$, moderate effect). Primary concerns showed no significant association with productivity at Q1, but at Q2 productivity perceptions differed by primary concern (Kruskal-Wallis; $\chi^2 = 18.34$, $p = 0.005$, $\varepsilon^2 = 0.19$, large effect), with those citing tool dependency reporting the highest productivity gains.


We next examined the developer experience--productivity relationship through Spearman correlations. All three developer experience dimensions showed significant positive correlations with productivity at both time points (Holm-Bonferroni corrected). At Q1, flow state was strongest ($\rho = 0.49$, $p_{adj} < 0.001$, moderate effect). By Q2, all three correlations weakened, and feedback loops replaced flow state as the strongest ($\rho = 0.37$, $p_{adj} < 0.001$, moderate effect), while flow state dropped to a small effect ($\rho = 0.20$, $p_{adj} < 0.05$).

\paragraph{Longitudinal Change in Productivity}

Next, we examined how productivity perceptions changed within participants ($n = 95$ matched participants). Paired Wilcoxon signed-rank tests showed no significant change ($\Delta M = -0.05$, $V = 99$, $p = 0.326$, $r_{rb} = -0.22$, small effect). Stability was high: 77\% maintained identical ratings, while 14\% decreased and 9\% increased intensity. No participant transitioned to negative perceptions.



\begin{figure}[htbp]
\centering
\includegraphics[width=1\linewidth]{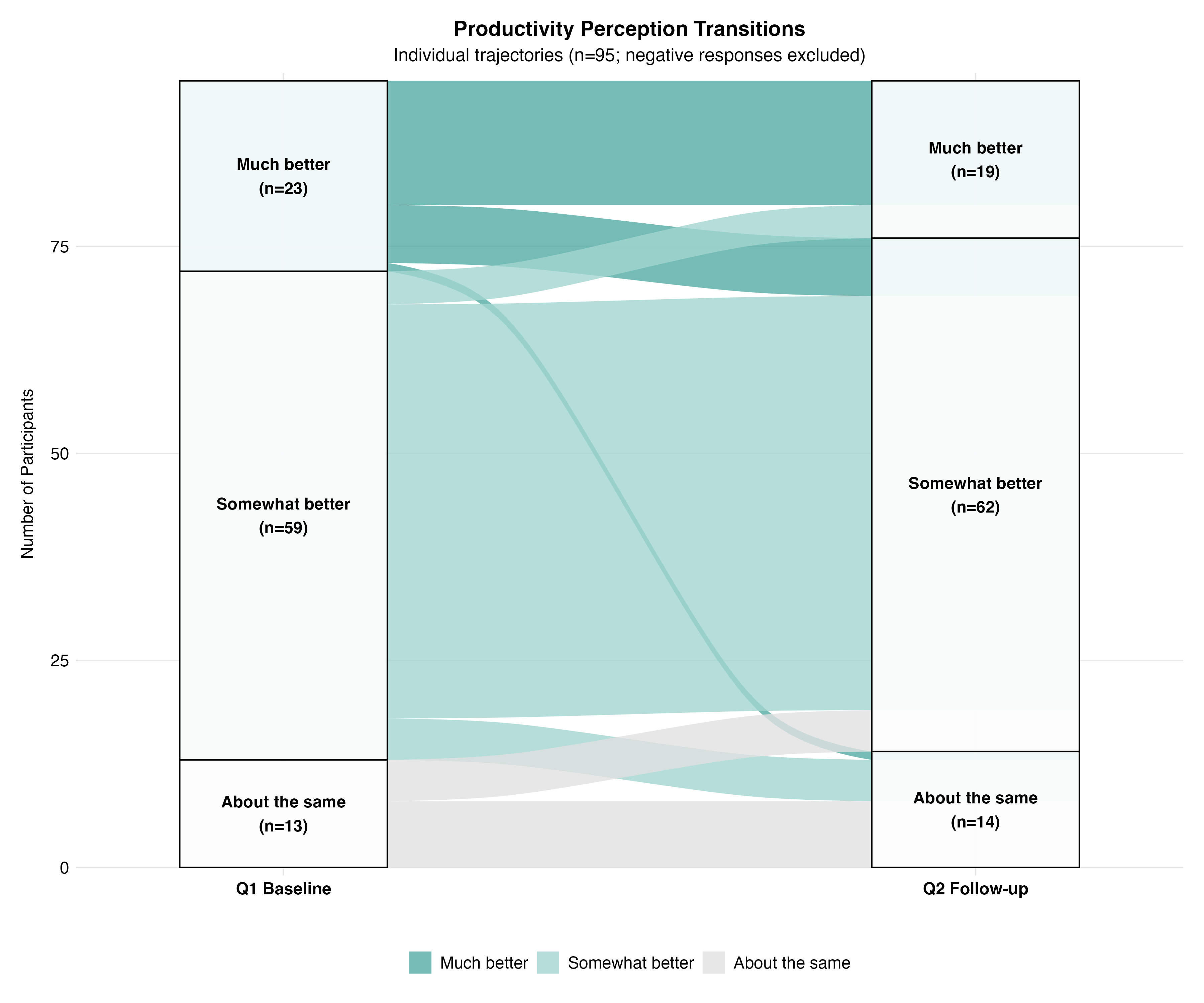}
\caption{Productivity perception transitions between Q1 and Q2.}
\Description{Chart showing transitions between productivity perceptions across the two time points. None of the participants shown rated productivity as worse at either time point. Most participants give the same rating at both time points, particularly those who rated productivity somewhat better at Q1. Of those who rated it much better at Q1, most stay much better and the rest mostly move to somewhat better. Of those who rated it about the same at Q1, most stay the same and some move to somewhat better.}
\label{fig:productivity-transitions}
\end{figure}

Stability varied by initial position (Figure~\ref{fig:productivity-transitions}): 85\% of \enquote{Somewhat better} respondents maintained their rating compared to 65\% of \enquote{Much better} and 62\% of neutral. Movement from neutral was entirely positive, with 38\% shifting to \enquote{Somewhat better}.

We correlated change scores (Q2 minus Q1) for each developer experience dimension with productivity change ($n = 94$, Holm-Bonferroni corrected). None reached significance (smallest $p_{adj} = 0.313$), contrasting with the moderate-to-strong cross-sectional correlations reported above.

\subsubsection{Qualitative Perspectives on Experience and Productivity}

Reflexive thematic analysis of responses to the questions asking about instances where AI had significantly helped or hindered work (Q1) or unexpected benefits or challenges (Q2), identified three themes capturing accelerated throughput with uneven experience, expanded capability in unfamiliar work, and verification as ongoing engineering work.

\paragraph{Theme 1: Accelerated throughput with uneven day-to-day experience.}
Across both time points, participants described AI coding assistants as accelerating progress, particularly for early-stage or repetitive work such as boilerplate generation, test scaffolding, and quick prototyping. Many framed this as a tangible reduction in effort, for example: \emph{\enquote{Boring mechanical tasks like refactoring files, updating tests, adding translations, creating test fixtures are 10x less effort with AI assistants}} (P132, Q1). Others described compressed delivery timelines: \emph{\enquote{It helped me complete the initial version of the project within hours instead of days \dots It made the entire process much faster and more efficient}} (P91, Q2). AI support was most valued when it enabled rapid starts, reduced mechanical effort, and sustained momentum. 

However, participants emphasised that these throughput gains were not uniformly experienced as smoother or easier work. Many described hidden costs in the form of repeated prompting, adaptation effort, or time lost to incorrect output. At Q1, some described prompt loops and rework, for example: \emph{\enquote{wasted some time getting into prompt loop}} (P18, Q1). Others reverted to traditional sources when AI was unreliable: \emph{\enquote{\dots it would reference methods which were no longer relevant, even after specifying a specific version I was using, I ended up going to docs}} (P26, Q1). By Q2, participants continued to describe speed alongside friction. Some described the variability directly: \emph{\enquote{Sometimes you gain productivity boost, the other time your wasting time on hallucinations or fixing the generated code}} (P89, Q2). Others noted that flow could be disrupted even as output accelerated, for example: \emph{\enquote{This does break the flow, but still speeds up the overall process}} (P28, Q2) and \emph{\enquote{less `in the zone' time, because while code is being produced, I now find myself switching context more often}} (P159, Q2). Altogether, AI often accelerated throughput while redistributing effort into supervision, correction, and verification. Perceived productivity and lived experience did not always move together.

\paragraph{Theme 2: Expanded perceived capability and willingness to engage with unfamiliar work.}
Participants described AI coding assistants as lowering the threshold for engaging with unfamiliar technologies, languages, and problem domains at both time points. Rather than claiming mastery, accounts emphasised practical reach, being able to start, explore, or contribute in areas that previously felt inaccessible or too time-consuming to approach. This was often described as a faster path from uncertainty to a workable first step, for example: \emph{\enquote{I don't have a lot of experience in authentication flows and got good directions from Gemini on designing a solution}} (P84, Q1) and \emph{\enquote{More confident picking up new tools that I'm otherwise unfamiliar with even if they don't have beginner friendly docs}} (P175, Q2). 

This expanded capability was framed as reducing hesitation and encouraging experimentation through readily available examples, explanations, and first-pass implementations. However, some cautioned that AI-enabled progress could short-circuit learning even while increasing output: \emph{\enquote{Destroys process of learning something new - you get a working solution, but not understanding why and how it may work}} (P41, Q1). Others cautioned critical judgement and active oversight were needed rather than passive trust: \emph{\enquote{I do like the fact that I can ask it about completely unfamiliar domains and it has a reasonable.. sounding at least.. response. I generally feel that I'm critical of all output, but am usually pleasantly surprised}} (P75, Q2). Overall, these accounts position AI coding assistants as expanding perceived capability by making unfamiliar work more approachable and actionable, though sometimes at the expense of deeper understanding. They also offer one reading of why perceived productivity held steady while developer experience ratings declined: the capability gains participants described did not diminish over the study period.

\paragraph{Theme 3: Verification and trust management as ongoing engineering work.}
AI coding assistants introduced new forms of verification and trust management that remained firmly within the engineer's remit. Participants framed correctness, quality, and consequence as fundamentally human responsibilities rather than treating AI output as authoritative. This required active review, judgement, and intervention. Trust calibration through practices such as sceptical oversight, deliberate steering, and collaborative verification within teams recurred across both time points. These accounts position trust management as an ongoing form of engineering work that shapes how AI-generated code is evaluated, integrated, and governed in practice.

\emph{Sub-theme 3A: Individual oversight: scepticism, accountability, and steering practices.}
Participants framed AI-generated output as provisional, emphasising that responsibility for correctness and consequences remained with the engineer. This accountability manifested through both scepticism and concrete steering practices. At Q1, participants articulated the need for active control, for example: \emph{\enquote{You need to always be fully in control, and guide the AI into the direction you want, and not let the AI guide you}} (P101, Q1), with some describing early over-trust as a corrective experience: \emph{\enquote{It's gotten code templates completely wrong (yaml) and I was overconfident in the output}} (P129, Q1). This scepticism translated into iterative refinement: \emph{\enquote{Often, it takes multiple tries and prompts to achieve the desired results}} (P3, Q1). Steering sometimes required more precise articulation of intent, for example \emph{\enquote{E.g. "Track the width of this react component" didn't work, but "Use a react reference to track the width of this react component when it's rendered" did work}} (P82, Q1). By Q2, participants described both recognising model failures and intervening accordingly, for example: \emph{\enquote{Identifying when the LLM has lost the thread of the problem and halting it is essential}} (P8, Q2). Constraint-setting became more explicit, such as deliberately breaking work into smaller units to reduce hallucination and drift: \emph{\enquote{it's been beneficial to break it down into smaller more specific "modules" and then tie them together at the end}} (P79, Q2). Across both time points, effective AI use was framed as requiring ongoing human judgement---through deliberate scepticism and active steering---rather than increasing reliance or unquestioned trust in AI output.

\emph{Sub-theme 3B: Collective trust, reputation, and collaborative implications.}
Participants described trust management as extending beyond individual practice into shared team contexts, where AI coding assistants affected code review effort, shared understanding, and reputational risk. At Q2, verification concerns extended to reputational and team-level stakes: \emph{\enquote{I might use AI chat to help with a question but I do not trust the answer enough for me to hang my reputation on it. I'll always look for a corroborating answer elsewhere}} (P76, Q2). These concerns extended to how participants judged others' AI-generated work, with some expressing reduced confidence in the author's understanding of the code: \emph{\enquote{If I know that a dev is using AI heavily I trust them much less to know what exactly their code is doing when reviewing their code}} (P152, Q2), alongside worries about unchecked exploration by less experienced colleagues, such as \emph{\enquote{When left unsupervised with AI, junior colleagues sometimes have a tendency to pursue inappropriate solutions to the wrong problems much further}} (P158, Q2). While these concerns were most explicit at Q2, Q1 accounts also signalled the need to retain human ownership over shared technical direction, for example: \emph{\enquote{Hence I keep control of the overall architecture}} (P55, Q1). Altogether, verification was positioned as both an individual and a shared supervisory responsibility, involving additional review discipline, corroboration practices, and clear accountability.

\subsubsection{Summary}
\label{subsec:results-rq2-summary}

Productivity perceptions were positive and stable, with 84\% reporting improvement at both time points. Developer experience showed cross-sectional associations with productivity: flow state showed the strongest association at Q1 ($\rho = 0.49$), though by Q2 feedback loops became the strongest. Among matched participants, 77\% gave identical productivity ratings over six months, while the proportion reporting worsened developer experience in at least one dimension grew from 14\% to 27\%. Feedback loops improved significantly, while cognitive load and flow state showed non-significant declines. Critically, changes in developer experience did not correlate with changes in productivity. Participants' accounts offer a possible interpretation of this divergence, describing accelerated throughput and expanded capability alongside hidden costs, with trust calibration emerging as ongoing engineering work.

\section{Discussion}
\label{chap:discussion}

This section interprets the findings presented in Section~\ref{chap:results} and 
discusses limitations, implications, and directions for future research.

\subsection{Task Focus Shifts and Emergence of Supervisory Engineering}
\label{sec:discussion-rq1}

Our results indicate that software engineers perceive AI coding assistants as reshaping where they direct their effort across development activities. Writing code showed the clearest decline, with 80\% and 82\% of participants reporting spending less time on this activity at each time point, respectively. This is consistent with prior research showing AI tools accelerate code generation~\cite{peng_impact_2023, cui_effects_2025}. Our qualitative findings help to explain where these perceived reductions in time came from. Participants described how AI handles boilerplate, scaffolding, and small fixes that previously consumed substantial effort (Theme 1, \emph{Compressing routine creation work}). They also described turning to AI before Google or Stack Overflow, folding what was once a separate information-gathering step into the coding interaction itself, as well as accelerating onboarding onto unfamiliar codebases, languages, and legacy code (Theme 2, \emph{Rerouting information seeking and comprehension into AI interactions}). Together, these patterns suggest that AI may compress not just the act of writing code but the preparatory work that surrounds it.

Studies using repository activity suggest similar shifts. Analysing open-source projects before and after Copilot's introduction, Xu et al.~\cite{xu_ai_2026} found that core developers reviewed 6.5\% more code while their original code productivity dropped by 19\%, measured from logged behaviour rather than self-report. Their setting differs from ours, and they report that the overall productivity increase was driven primarily by less experienced developers, so the convergence lies in the direction of the change rather than in its cause.


Some patterns ran counter to expectations. Given that previous research has found that verification consumes substantial time in AI-assisted development~\cite{mozannar_reading_2024}, we expected more participants to report spending more time on verification activities like testing and code review. Instead, reviewing code was the only task above neutral at either time point, and only modestly so. Testing remained below neutral throughout, meaning more engineers perceived spending less time on it even in an AI-assisted context. One potential explanation is that engineers do not always recognise the verification they perform on AI-generated output, such as evaluating whether to accept a suggestion, scanning output for errors, or checking that generated code integrates correctly, as \enquote{testing} or \enquote{code review} in the traditional sense. As a result, standard Software Development Life Cycle (SDLC) categories may not fully capture where this verification effort is actually going. 

Our qualitative findings suggest where some of this effort goes. We propose \emph{supervisory engineering work} (Theme 3, \emph{From doing to supervising AI-generated code}) as an interpretive construct offered for further investigation. It encompasses:
\begin{itemize}
    \item \textbf{Directing}: specifying intent, crafting prompts, providing context, and iterating when output misses the mark
    \item \textbf{Evaluating}: reading AI output and deciding what to accept, modify, or reject
    \item \textbf{Correcting}: fixing errors, integrating output into existing code, and maintaining consistency
\end{itemize}

Understood this way, supervisory engineering work is broader in kind than the review overhead already reported. That overhead maps onto evaluating and correcting: prior studies describe the effort of assessing AI-generated output~\cite{bird_taking_2022} and the time taken to review and fix it~\cite{vaillant_developers_2024,banh_copiloting_2025}. Even there the fit is imperfect, since evaluating sits inside authoring rather than after it, with the engineer judging output as it appears in code they are still writing rather than a finished contribution from another author. Directing lies outside it altogether, occurring prior to there being any output to assess and steering what the tool produces rather than judging what it has produced. What we propose is therefore a category that includes review overhead rather than a restatement of it.

Research on the interaction between people and automated systems has examined supervision in other settings. Writing about undersea teleoperators, Sheridan and Verplank~\cite{sheridan_human_1978} introduced the term \emph{supervisory control} for a person directing and monitoring a system that acts on its own between interventions. Parasuraman et al.~\cite{parasuraman_model_2000} later extended this concept across the stages of gathering information, analysing it, choosing a course of action, and carrying it out, and argued that automation does not simply remove human work but changes it, creating new coordination demands. This research also documented what the shift cost the people in it. Endsley and Kiris~\cite{endsley_out_1995} found that operators were slower to take over when an automated decision aid failed, which they attributed to the operator moving from actively working on the problem to passively watching it. They also found that the automation did not reduce the workload participants reported. Bainbridge~\cite{bainbridge_ironies_1983} put it more bluntly, describing a job reduced to monitoring as \enquote{very boring but very responsible} and difficult for those in it to come to terms with.

Our construct emerged from our own data in an AI-assisted software engineering setting, which differs in three respects. First, the system being supervised is generative: its output is not measured against a target state the way an industrial process is, so evaluating correctness has to be done on a case by case basis. Second, the supervisory work is itself active rather than the passive monitoring that much of this research treats as the source of these costs. Third, who does what is decided by the engineer, prompt by prompt, within a single task, rather than by a designer or by a scheme that varies with operating conditions~\cite{parasuraman_model_2000}. 

Related work has begun to make this connection: Simkute et al.~\cite{simkute_ironies_2025} revisit Bainbridge's ironies for generative AI in knowledge work, and, more recently, Dhanorkar et al.~\cite{dhanorkar_human_2026} examine oversight practices around autonomous agents. What we propose is more specific: a candidate category of work that emerged from how software engineers described their task focus shifting over the study period.

Since who does what is the engineer's to decide, not all engineers experience this supervisory work equally. Theme 4 from our thematic analysis, \emph{Verification and trust calibration moderating perceived benefits}, captured how engineers vary in their willingness to accept AI output, with some restricting AI to low-risk contexts while others accept suggestions more readily. This calibration shapes how much supervision is performed.

While the quantitative analysis showed a shift in the perceived balance from creation toward verification, the qualitative themes suggested less of a change between time points. The same themes were evident at both, with some participants placing more or less emphasis on supervision and verification at Q2 (Theme 3, \emph{From doing to supervising AI-generated code}; Theme 4, \emph{Verification and trust calibration moderating perceived benefits}), but no themes appearing or disappearing. One interpretation is that practices evolved gradually over the six months, intensifying existing ways of working rather than replacing them. Furthermore, since the open-ended workflow question at Q2 made no reference to Q1, responses may also have contrasted working with AI against working without it.


More broadly, these findings suggest that AI coding assistants may be reshaping the nature of software development work itself. If engineers spend less time writing code and more time directing and evaluating AI output, this has implications for which skills remain essential, what engineers practice day-to-day, and how the profession develops its next generation. 

\subsection{Developer Experience, Productivity, and Emergence of Productivity-Experience Paradox}
\label{sec:discussion-rq2}


Our results found that software engineers predominantly perceived AI coding assistants as improving their productivity, with 84\% reporting improvement at both time points, consistent with industry surveys~\cite{stackoverflow_ai_2024}. Perceived impact on developer experience, using the three DevEx dimensions of cognitive load, feedback loops, and flow state, was also generally positive at Q1, though less uniformly so: 66\% reported improved cognitive load, 62\% improved feedback loops, and 53\% improved flow state. Unlike productivity, developer experience showed concerning signs of erosion over the six-month study period, with the number of matched participants reporting worsened perceptions in at least one dimension increasing from 14\% in Q1 to 27\%. Our qualitative Theme 1, \emph{Accelerated throughput with uneven day-to-day experience}, helps explain this pattern: participants described impressive time compression alongside persistent friction from repeated prompting and correction.



Flow state was perceived as most vulnerable to erosion. Participants described features of AI-assisted workflows as potentially disruptive to flow. Each AI suggestion requires evaluation; each generation requires verification. The cycle of prompting, reviewing, accepting or rejecting, and iterating can constitute ongoing interruption built into the workflow itself. Our qualitative Theme 3, \emph{Verification and trust management as ongoing engineering work}, captured this dynamic: participants described needing to maintain vigilance when reviewing AI output, a form of mental effort that persists even when the code itself is generated quickly. Prior research found that context switching and interruption frequency are key factors affecting flow~\cite{razzaq_systematic_2024}. The delegation model may allow developers to turn attention elsewhere while code is being generated, but this shifts work from sustained focus to a rhythm of directing, waiting, and evaluating---precisely the supervisory engineering work we propose is emerging.




Our findings suggest that the established relationships between productivity and developer experience may need re-examination when AI is introduced. We term this pattern the \emph{productivity-experience paradox}: consistently positive productivity perceptions alongside eroding developer experience, potentially decoupling the established relationship between the two. Prior research found that uninterrupted focus is central to developers' sense of productivity~\cite{meyer_software_2014} and that satisfaction and productivity are bidirectionally linked~\cite{storey_towards_2021}. Our qualitative findings help explain why productivity held steady while developer experience lowered. Theme 1, \emph{Accelerated throughput with uneven day-to-day experience}, captured participants describing faster output alongside persistent friction, with perceived productivity and lived experience not always moving together. The DevEx framework treats its three dimensions as complementary drivers of productivity, but AI assistance may reshape their relative contributions, with faster feedback compensating for increased cognitive friction. Our correlations show this directly: the correlation between each dimension and perceived productivity weakened between time points, and flow state, the strongest at Q1 ($\rho = 0.49$), was displaced by feedback loops by Q2 while falling to a small effect ($\rho = 0.20$). Changes in the dimensions did not track changes in productivity at all, with none of the change-score correlations reaching significance. Whether this reconfiguration is sustainable, or whether the erosion in flow and cognitive load eventually undermines the gains, remains an open question.


This paradox describes coexisting perceptions rather than a causal trade-off. The two are parallel self-reports, and our design cannot establish whether one drives the other. The finding matters for what it shows about engineers' lived experience of AI-assisted work, not for any causal mechanism.

\subsection{The Shifting Tool Landscape}
\label{sec:synthesis-tools}

Our study captured a six-month window in a field evolving at unprecedented speed. The tool landscape shifted substantially even during data collection, with 82\% of matched participants changing their tool combinations between time points. 
Two patterns emerged. First, \emph{tool diversification}: participants mentioned around 37 distinct tools across the study, and engineers are not settling on a single tool but building toolkits. The average number of tools per engineer increased from 1.9 at Q1 to 2.9 at Q2. Second, \emph{shift toward coding-specialised tools}: across our full samples, ChatGPT declined from 70\% to 58\% usage, while Cursor grew from 16\% to 29\%. This suggests some engineers are moving from general-purpose conversational AI toward purpose-built coding environments.

The only concern that changed significantly over the six-month study period was maintainability, which increased. This may reflect growing familiarity with AI-generated code, and a recognition that AI optimises for \emph{works now} rather than \emph{maintains later}. One participant echoed this, listing \emph{\enquote{is it maintainable}} (P88, Q2) among the checks contributing to their verification time (Theme 4, \emph{Verification and trust calibration moderating perceived benefits}).

\subsection{Threats to Validity}
\label{sec:threats-to-validity}

Several threats to validity warrant attention when interpreting our findings. First, regarding internal validity, 40\% of participants who completed Q1 did not complete Q2, reducing our matched longitudinal sample from 158 to 95. As reported in Section~\ref{chap:methodology}, demographic comparisons showed no significant differences between retained and lost participants, supporting the representativeness of the matched cohort. However, we cannot rule out unmeasured differences in attitudes or experiences. Engineers who became disillusioned with AI tools may have been less motivated to complete a follow-up questionnaire about them, which would cause our longitudinal findings to overestimate stability and underestimate negative trajectories. We did not conduct a formal a priori power analysis because we relied on convenience and referral-chain sampling, and the sample size represents the participants who were available during the study period. Non-significant longitudinal results should therefore not be read as evidence of absence: small true effects may have gone undetected.

As disclosed in Section~\ref{chap:methodology}, the first author is a practising software engineer whose familiarity with the domain may have influenced the framing of questions, interpretation of findings, or emphasis on certain themes. We managed this threat to internal validity through iterative discussions with the research supervisor throughout data collection and analysis, including survey design and interpretation of both quantitative patterns and qualitative themes. However, this potential bias cannot be completely eliminated.

Threats to external validity include limitations of our sample. Our study only included engineers currently using AI coding assistants; those who tried these tools and abandoned them are absent from our sample. This selection bias likely causes our findings to overestimate positive outcomes, as engineers who found AI unhelpful would have stopped using it and thus would not appear in our data. The 84\% positive productivity rate should be interpreted as \enquote{84\% of continuing users}, not \enquote{84\% of all who tried}. Beyond this survivor bias, 85\% of our sample were men and predominantly English-speaking, which limits generalisability to under-represented groups. Participation was also geographically concentrated, with New Zealand accounting for 41\% of participants at Q1 and 48\% at Q2. The sample is therefore not statistically representative of the global software engineering population, and both self-selection into the study and the platforms we recruited through are likely to have influenced who took part. Finally, some groups were too small to compare separately. For example, only around 2\% of participants reported spending more time writing code, too few to characterise as a group, and future work could sample such engineers deliberately.

Another external validity threat is temporal. Our findings reflect experiences with AI coding assistants as they existed in late 2024 and early 2025. Rapid tool evolution means patterns may not hold with newer agentic tools that operate more autonomously. In addition, our data is confounded with a rapidly changing tool landscape; our findings reflect practice-in-the-wild rather than the maturation of any single tool. Changes we observed may stem from tool switching, tool improvement, or growing familiarity, which we cannot fully disentangle; we did not capture how long participants had been using AI coding assistants before the study. Organisational context is a further influence we could not assess, since we did not ask about employer policies or practices around AI use, which may themselves have changed over the study period.

A threat to construct validity is our use of perception-based measures. We captured perceptions rather than objective performance. As discussed in Section~\ref{chap:literature-review}, perceived and actual productivity can diverge substantially, and several factors may inflate positive perceptions in our data: social desirability, effort justification, and question framing. However, our interpretivist stance treats perceptions as meaningful data in their own right: they drive adoption decisions, shape professional identity, and influence how engineers approach their work.

Single-item measurement poses a related threat to construct validity. With one indicator per construct we cannot estimate measurement error, so small differences between tasks or between time points should be interpreted cautiously.

Participants' understanding of the six development tasks is a further threat to construct validity. As stated in Section~\ref{subsec:questionnaire-structure}, they were presented as labels without definition. Each rating therefore reflects what that task meant to that participant rather than a definition we supplied.

Recall bias is another threat to construct validity. Engineers may overweight memorable events when summarising months of AI use, while recency effects may inflate the influence of experiences closest to each questionnaire. Internal benchmarks may also shift: an engineer who found AI frustrating six months ago may have recalibrated what counts as frustrating, making direct comparisons across time points imprecise. From our interpretivist perspective, however, these distortions are not merely noise to be eliminated but part of what we sought to understand.

Finally, our convergent parallel design limits how far the quantitative and qualitative results can explain one another. Both were collected in the same questionnaire at each time point, so we could not observe a quantitative pattern and then return to participants to probe what lay behind it. Explanatory depth of that kind would require a sequential design, with the qualitative work following the quantitative analysis.

\subsection{Implications}
\label{sec:implications}

\subsubsection{For Individual Engineers} The proposed category of supervisory engineering work suggests preparing for a shift in daily practice, with less time writing code and more time directing, evaluating, and correcting AI output. This may be less freed-up time than differently allocated effort. Verification and trust calibration are important to get the most benefits from AI. Knowing when to accept and when to reject AI output emerged across our qualitative findings as an important skill, and one that transfers across the rapidly evolving tool landscape.

\subsubsection{For Organisations and Teams} Organisations should recognise that engineers' effort may be shifting rather than simply reducing, from creation toward verification and toward the supervisory engineering work we propose. 
The perceived effort reallocation has implications for how organisations structure teams, evaluate performance, and prioritise skills when hiring, particularly as verification, judgement, and trust calibration become more central to the role. At the team level, participants described trusting a colleague's understanding of their own code less when they knew AI had done much of the work, and noted junior engineers pursuing unsuitable solutions further when unsupervised. Making AI use visible, adjusting review practices accordingly, and overseeing junior engineers' AI work more deliberately may all help. On measurement, output metrics alone may mask experiential erosion, so monitoring developer experience alongside productivity could provide earlier warning of unsustainable patterns. On tooling, enabling experimentation rather than mandating a single tool may better serve adaptation, given how substantially the landscape shifted even during our six-month study.

\subsubsection{For Educators} If AI is shifting engineering work from creation toward verification while introducing new responsibilities like supervisory engineering work, educational programmes may need to prepare students accordingly. This means teaching not just traditional technical skills but also judgement, trust calibration, and the components of supervisory work: directing, evaluating, and correcting AI output.  More broadly, transferable principles of human-AI collaboration may prove more durable than tool-specific training given how rapidly the landscape is evolving. Assessment methods may also need to evolve, evaluating understanding rather than output quality alone, since students can now produce working code without necessarily understanding it.

\subsubsection{For Tool Builders} As the perceived balance of our participants' work shifted toward verification, so too should the focus of the tools they use. The supervisory engineering work we propose would introduce further requirements, since directing, evaluating and correcting sit outside the traditional development activities and are therefore what today's tools support least well. Interfaces for steering would help with directing, particularly where intent has to be restated before the output is right. In terms of evaluating, engineers would benefit from seeing why a suggestion was produced, as well as being able to tell whether it works and why, since verifying code the engineer did not write is cognitively heavy. Tools could also do more for correcting, helping engineers integrate generated code into the existing solution rather than treating a suggestion as complete once accepted. The agentic tools that have emerged since our study would place even further demands on tools, since an engineer cannot direct, evaluate, or correct what they cannot see: understanding what such a tool is doing, how far it has progressed, and when it has drifted from the problem matter as much as the quality of any individual suggestion. On developer experience, only feedback loops improved significantly, so fast iteration appears well served; cognitive load and flow state did not. Flow state was the predominant issue in the Negative cohort, at 76\% by Q2, which suggests a need to focus on protecting concentration and reducing interruption as core features of the tools engineers use when they build with AI.


\subsection{Future Research Directions}
\label{sec:future-research}

\subsubsection{Emerging Competencies and Work Categories.} AI assistance may be introducing new work categories that existing frameworks do not capture. The supervisory engineering work we proposed raises questions about whether this represents a genuinely new SDLC activity distinct from traditional categories like designing, coding, testing, and reviewing. Future research could examine whether supervisory engineering work is empirically distinct from existing activities, what skills it requires, and how engineers should develop them. If this work category proves real and distinct, established SDLC models may need updating to account for where developer effort now goes. That effort may also be distributed differently between engineers. Recent research has found that students with different cognitive styles engage with AI coding assistants in different ways, including differences in validation behaviour, trust formation, and willingness to rely on AI-generated solutions~\cite{cognitive_inclusivity_copilot}. Future research could examine whether professional engineers similarly adopt distinct work styles when supervising AI, for example placing more emphasis on directing, evaluating or correcting, and what factors shape those styles.

\subsubsection{Developer Experience and Productivity.} Prior research has shown that developer experience drives productivity, yet our data suggests a productivity-experience paradox, with productivity holding steady while two of the three DevEx dimensions worsened. Future research could examine whether the DevEx dimensions contribute differently in AI-assisted contexts, whether new dimensions are needed, and whether the relationship between developer experience and productivity has changed. Longer-term studies tracking both productivity and experience could illuminate whether the divergence we observed is sustainable over longer time horizons or whether worsening flow state and cognitive load manifest as burnout or turnover.

\subsubsection{Satisfaction in Supervisory Work.} The shift toward supervisory engineering work raises questions about developer satisfaction. The worsening flow state and cognitive load we observed suggest some experiential costs may accompany this transition. Future research could examine whether engineers find supervisory work intrinsically satisfying or whether reduced hands-on coding diminishes professional fulfilment and whether these effects vary across developers. If supervisory work proves less satisfying, interventions could be devised to improve satisfaction. 

\subsubsection{Team and Organisational Dynamics.} Our study focused on individual engineers, but the creation-to-verification shift and supervisory engineering work we described likely have team-level implications. 
Research examining team dynamics and organisational structures could illuminate how individual-level findings aggregate at the team level and whether supervisory engineering and verification efforts are equally distributed.

\section{Conclusion}
\label{chap:conclusion}

This study examined how professional software engineers perceive the effects of AI coding assistants on their work, tracking those perceptions over time. Using a longitudinal mixed-methods design with questionnaires administered at two time points six months apart allowed us to see shifts that cross-sectional snapshots would miss.

Our findings suggest that development work is being reorganised rather than simply accelerated. Most engineers perceived spending less time across traditional development tasks, with the relative balance shifting from creation toward verification; we have termed this the \emph{creation-to-verification shift}. Yet the increase in verification was modest, raising the question of where the effort is going. We propose this effort flows into what we have called \emph{supervisory engineering work}: directing AI, evaluating its output, and correcting its errors. Meanwhile, the \emph{productivity-experience paradox} we documented suggests these tools may be decoupling dimensions that previously moved together, with productivity perceptions remaining consistently positive even as developer experience showed signs of erosion over time, particularly in flow state and cognitive load. Together, these patterns suggest that AI coding assistants may be reshaping the nature of software engineering work, not merely making existing tasks faster.


Whether AI coding assistants will finally close the persistent gap between ambition and achievement that has defined software engineering since its origins remains to be seen. Brooks warned that technology alone has historically failed to deliver dramatic improvements~\cite{brooks_no_1987}, and our findings suggest that unintended consequences may be emerging. As software engineering serves as a proving ground for AI-assisted work, these patterns may foreshadow dynamics that other professions will encounter as AI tools spread. For a global community of professionals who have built careers around the craft of building software, understanding how engineers perceive these tools is essential, because perceptions drive behaviour. This study contributes to that understanding, informing educators preparing the next generation, organisations supporting their teams, those building the tools themselves, and engineers navigating their own paths through this period of rapid change.

\section*{Replication Package}
The replication package for this study is openly available~\cite{replication}. It contains the full questionnaire instrument administered at both time points, the codebook developed during the qualitative analysis, the R packages used for the quantitative analysis, and the full list of AI coding assistants participants reported using. Readers can use this to see exactly what was asked of participants, how their open-ended responses were coded, and which tools lie behind the figures reported in Section~\ref{chap:results}. Individual responses are not included, because participants did not consent to their responses being shared.
\section*{Use of GenAI}
We used Claude Code (Anthropic) to assist with developing R scripts for statistical analysis and reviewing written text for clarity and readability. All outputs were critically reviewed and validated, and responsibility for the correctness of the analysis and interpretations rests entirely with the researchers.

\section*{Acknowledgments}
We thank all participants for their time and input into this research. This research was partially supported by Westpac New Zealand, Annie Vella's employer at the time the research was conducted, and a Rutherford Discovery Fellowship administered by the Royal Society Te Ap\=arangi, New Zealand.

\bibliographystyle{ACM-Reference-Format}
\bibliography{references}

@misc{replication,
    title = {Replication package for ``The Impact of AI Coding Assistants on Software Engineering: A Longitudinal Study''},
    author = {Annie Vella and Kelly Blincoe},
    year = {2026},
    url={https://doi.org/10.5281/zenodo.21896822}
}

@inproceedings{alami_human_2025,
	title = {Human and Machine: How Software Engineers Perceive and Engage with {AI}-Assisted Code Reviews Compared to Their Peers},
	url = {https://ieeexplore.ieee.org/document/11024268},
	doi = {10.1109/CHASE66643.2025.00016},
	booktitle = {2025 {IEEE}/{ACM} 18th International Conference on Cooperative and Human Aspects of Software Engineering ({CHASE})},
	author = {Alami, Adam and Ernst, Neil},
	month = apr,
	year = {2025},
	pages = {63--74},
    publisher={IEEE},
    address = {New York, NY, USA},
}

@article{baltes_sampling_2022,
	title = {Sampling in software engineering research: a critical review and guidelines},
	volume = {27},
	shorttitle = {Sampling in software engineering research},
	url = {https://doi.org/10.1007/s10664-021-10072-8},
	doi = {10.1007/s10664-021-10072-8},
	number = {94},
	journal = {Empirical Software Engineering},
	author = {Baltes, Sebastian and Ralph, Paul},
	month = apr,
	year = {2022},
    pages={1--31},
}

@misc{becker_measuring_2025,
	title = {Measuring the Impact of Early-2025 {AI} on Experienced Open-Source Developer Productivity},
	url = {https://metr.org/blog/2025-07-10-early-2025-ai-experienced-os-dev-study/},
	author = {Becker, Joel and Rush, Nate and Barnes, Beth and Rein, David},
	year = {2025},
	howpublished = {METR},
}

@article{bird_taking_2022,
	title = {Taking Flight with Copilot: Early insights and opportunities of {AI}-powered pair-programming tools},
	volume = {20},
	shorttitle = {Taking Flight with Copilot},
	url = {https://dl.acm.org/doi/10.1145/3582083},
	doi = {10.1145/3582083},
	number = {6},
	journal = {Queue},
	author = {Bird, Christian and Ford, Denae and Zimmermann, Thomas and Forsgren, Nicole and Kalliamvakou, Eirini and Lowdermilk, Travis and Gazit, Idan},
	month = dec,
	year = {2022},
	pages = {35--57},
}

@article{braun_reflecting_2019,
	title = {Reflecting on reflexive thematic analysis},
	volume = {11},
	url = {https://www.tandfonline.com/doi/full/10.1080/2159676X.2019.1628806},
	doi = {10.1080/2159676x.2019.1628806},
	number = {4},
	journal = {Qualitative Research in Sport, Exercise and Health},
	author = {Braun, Virginia and Clarke, Victoria},
	month = aug,
	year = {2019},
	pages = {589--597},
}

@article{braun_using_2006,
	title = {Using thematic analysis in psychology},
	volume = {3},
	url = {http://www.tandfonline.com/doi/abs/10.1191/1478088706qp063oa},
	doi = {10.1191/1478088706qp063oa},
	number = {2},
	journal = {Qualitative Research in Psychology},
	author = {Braun, Virginia and Clarke, Victoria},
	month = jan,
	year = {2006},
	pages = {77--101},
}

@article{brooks_no_1987,
	title = {No {Silver} {Bullet} {Essence} and {Accidents} of {Software} {Engineering}},
	volume = {20},
	copyright = {https://ieeexplore.ieee.org/Xplorehelp/downloads/license-information/IEEE.html},
	url = {http://ieeexplore.ieee.org/document/1663532/},
	doi = {10.1109/mc.1987.1663532},
	number = {4},
	journal = {Computer},
	author = {{Brooks}},
	month = apr,
	year = {1987},
	pages = {10--19},
}

@inproceedings{butler_dear_2025,
	title = {Dear {Diary}: {A} {Randomized} {Controlled} {Trial} of {Generative} {AI} {Coding} {Tools} in the {Workplace}},
	shorttitle = {Dear {Diary}},
	url = {https://ieeexplore.ieee.org/abstract/document/11121735},
	doi = {10.1109/ICSE-SEIP66354.2025.00034},
	booktitle = {2025 {IEEE}/{ACM} 47th {International} {Conference} on {Software} {Engineering}: {Software} {Engineering} in {Practice} ({ICSE}-{SEIP})},
	author = {Butler, Jenna and Suh, Jina and Haniyur, Sankeerti and Hadley, Constance},
	month = apr,
	year = {2025},
	pages = {319--329},
    publisher={IEEE},
    address = {New York, NY, USA},
}

@misc{chatterjee_impact_2024,
	title = {The Impact of {AI} Tool on Engineering at {ANZ} Bank: An Empirical Study on {GitHub} {Copilot} within Corporate Environment},
	shorttitle = {The Impact of {AI} Tool on Engineering at {ANZ} Bank},
	url = {http://arxiv.org/abs/2402.05636},
	doi = {10.48550/arXiv.2402.05636},
	publisher = {arXiv},
	author = {Chatterjee, Sayan and Liu, Ching Louis and Rowland, Gareth and Hogarth, Tim},
	month = feb,
	year = {2024},
}

@inproceedings{chen_impact_2024,
	title = {The Impact of {AI}-Pair Programmers on Code Quality and Developer Satisfaction: Evidence from {TiMi} studio},
	url = {https://dl.acm.org/doi/10.1145/3665348.3665383},
	doi = {10.1145/3665348.3665383},
	booktitle = {2024 International Conference on Generative Artificial Intelligence and Information Security ({GAIIS})},
	author = {Chen, Tianyi},
	month = may,
	year = {2024},
	pages = {201--205},
    publisher={ACM},
    address={New York, NY, USA}
}

@book{creswell_designing_2018,
	address = {Thousand Oaks, California, USA},
	edition = {Third},
	title = {Designing and {Conducting} {Mixed} {Methods} {Research}},
	publisher = {SAGE},
	author = {Creswell, John W. and Plano Clark, Vicki L.},
	year = {2018},
}

@misc{cui_effects_2025,
	address = {Rochester, NY, USA},
	type = {{SSRN} {Scholarly} {Paper}},
	title = {The Effects of Generative {AI} on High-Skilled Work: Evidence from Three Field Experiments with Software Developers},
	shorttitle = {The Effects of Generative {AI} on High-Skilled Work},
	url = {https://papers.ssrn.com/abstract=4945566},
	doi = {10.2139/ssrn.4945566},
	publisher = {Social Science Research Network},
	author = {Cui, Zheyuan (Kevin) and Demirer, Mert and Jaffe, Sonia and Musolff, Leon and Peng, Sida and Salz, Tobias},
	month = feb,
	year = {2025},
}

@article{dangelo_what_2024,
	title = {What {Do} {Developers} {Want} {From} {AI}?},
	volume = {41},
	url = {https://ieeexplore.ieee.org/document/10493171},
	doi = {10.1109/MS.2024.3363538},
	number = {3},
	journal = {IEEE Software},
	author = {D'Angelo, Sarah and Murillo, Ambar and Chandra, Satish and Macvean, Andrew},
	month = may,
	year = {2024},
	pages = {11--15},
}

@article{depalma_exploring_2024,
	title = {Exploring {ChatGPT}'s code refactoring capabilities: {An} empirical study},
	volume = {249},
	url = {https://www.sciencedirect.com/science/article/pii/S0957417424004676},
	doi = {10.1016/j.eswa.2024.123602},
	journal = {Expert Systems with Applications},
	author = {DePalma, Kayla and Miminoshvili, Izabel and Henselder, Chiara and Moss, Kate and AlOmar, Eman Abdullah},
	month = sep,
	year = {2024},
	pages = {123602},
}

@techreport{dora_ai_2025,
	title = {State of {AI}-assisted Software Development},
	url = {https://dora.dev/research/2025/dora-report/},
	institution = {Google},
	author = {{DORA}},
	year = {2025},
}

@misc{hoffmann_generative_2025,
	address = {Rochester, NY, USA},
	type = {{SSRN} {Scholarly} {Paper}},
	title = {Generative {AI} and the {Nature} of {Work}},
	url = {https://papers.ssrn.com/abstract=5007084},
	doi = {10.2139/ssrn.5007084},
	publisher = {Social Science Research Network},
	author = {Hoffmann, Manuel and Boysel, Sam and Nagle, Frank and Peng, Sida and Xu, Kevin},
	month = apr,
	year = {2025},
}

@misc{houck_space_2025,
	title = {The {SPACE} of {AI}: {Real}-{World} {Lessons} on {AI}'s {Impact} on {Developers}},
	shorttitle = {The {SPACE} of {AI}},
	url = {http://arxiv.org/abs/2508.00178},
	doi = {10.48550/arXiv.2508.00178},
	publisher = {arXiv},
	author = {Houck, Brian and Lowdermilk, Travis and Beyer, Cody and Clarke, Steven and Hanrahan, Ben},
	month = jul,
	year = {2025},
}

@article{huang_impact_2025,
	title = {The impact of {GenAI}-enabled coding hints on students' programming performance and cognitive load in an {SRL}-based {Python} course},
	volume = {56},
	url = {https://onlinelibrary.wiley.com/doi/abs/10.1111/bjet.13589},
	doi = {10.1111/bjet.13589},
	number = {5},
	journal = {British Journal of Educational Technology},
	author = {Huang, Anna Y. Q. and Lin, Cheng-Yan and Su, Sheng-Yi and Yang, Stephen J. H.},
	year = {2025},
	pages = {1942--1972},
}

@article{inman_seamful_2025,
	title = {Seamful {AI} for Creative Software Engineering},
	volume = {42},
	url = {https://ieeexplore.ieee.org/document/10857384},
	doi = {10.1109/MS.2025.3534085},
	journal = {IEEE Software},
	author = {Inman, Sarah and Murillo, Ambar and D'Angelo, Sarah and Brown, Adam and Green, Collin},
	year = {2025},
    pages={59--66},
}

@misc{jimenez_swe_2023,
	title = {{SWE}-bench: {Can} {Language} {Models} {Resolve} {Real}-{World} {GitHub} {Issues}?},
	shorttitle = {{SWE}-bench},
	url = {http://arxiv.org/abs/2310.06770},
	doi = {10.48550/arXiv.2310.06770},
	publisher = {arXiv},
	author = {Jimenez, Carlos E. and Yang, John and Wettig, Alexander and Yao, Shunyu and Pei, Kexin and Press, Ofir and Narasimhan, Karthik},
	month = oct,
	year = {2023},
}

@article{kuhail_will_2024,
	title = {"{Will} {I} be replaced?" {Assessing} {ChatGPT}'s effect on software development and programmer perceptions of {AI} tools},
	volume = {235},
	shorttitle = {"{Will} {I} be replaced?},
	url = {https://www.sciencedirect.com/science/article/pii/S0167642324000340},
	doi = {10.1016/j.scico.2024.103111},
	journal = {Science of Computer Programming},
	author = {Kuhail, Mohammad Amin and Mathew, Sujith Samuel and Khalil, Ashraf and Berengueres, Jose and Shah, Syed Jawad Hussain},
	month = jul,
	year = {2024},
	pages = {103111},
}

@misc{kumar_intuition_2025,
	title = {Intuition to {Evidence}: {Measuring} {AI}'s {True} {Impact} on {Developer} {Productivity}},
	shorttitle = {Intuition to {Evidence}},
	url = {http://arxiv.org/abs/2509.19708},
	doi = {10.48550/arXiv.2509.19708},
	publisher = {arXiv},
	author = {Kumar, Anand and Khare, Vishal and Sharma, Deepak and Kumar, Satyam and Saini, Vijay and Yadav, Anshul and Jain, Sachendra and Rana, Ankit and Verma, Pratham and Meena, Vaibhav and Edubilli, Avinash},
	month = sep,
	year = {2025},
}

@inproceedings{lange_exploring_2025,
	address = {Cham},
	title = {Exploring {Flow} in {IT} {Professionals}' {Use} of {AI}-{Integrated} {Tools}: {Insights} from {Interviews}},
	shorttitle = {Exploring {Flow} in {IT} {Professionals}' {Use} of {AI}-{Integrated} {Tools}},
	doi = {10.1007/978-3-031-93429-2_3},
	booktitle = {Artificial {Intelligence} in {HCI}},
	publisher = {Springer Nature Switzerland},
	author = {Lange, Eve Martina and Cajander, Åsa and Normark, Maria},
	editor = {Degen, Helmut and Ntoa, Stavroula},
	year = {2025},
	pages = {44--58},
}

@inproceedings{liang_large_2024,
	address = {New York, NY, USA},
	series = {{ICSE} '24},
	title = {A {Large}-{Scale} {Survey} on the {Usability} of {AI} {Programming} {Assistants}: {Successes} and {Challenges}},
	shorttitle = {A {Large}-{Scale} {Survey} on the {Usability} of {AI} {Programming} {Assistants}},
	url = {https://dl.acm.org/doi/10.1145/3597503.3608128},
	doi = {10.1145/3597503.3608128},
	booktitle = {Proceedings of the {IEEE}/{ACM} 46th {International} {Conference} on {Software} {Engineering}},
	publisher = {Association for Computing Machinery},
	author = {Liang, Jenny T. and Yang, Chenyang and Myers, Brad A.},
	month = feb,
	year = {2024},
	pages = {1--13},
}

@inproceedings{meyer_software_2014,
	address = {New York, NY, USA},
	series = {{FSE} 2014},
	title = {Software developers' perceptions of productivity},
	url = {https://doi.org/10.1145/2635868.2635892},
	doi = {10.1145/2635868.2635892},
	booktitle = {Proceedings of the 22nd {ACM} {SIGSOFT} {International} {Symposium} on {Foundations} of {Software} {Engineering}},
	publisher = {Association for Computing Machinery},
	author = {Meyer, André N. and Fritz, Thomas and Murphy, Gail C. and Zimmermann, Thomas},
	month = nov,
	year = {2014},
	pages = {19--29},
}

@article{meyer_work_2017,
	title = {The Work Life of Developers: Activities, Switches, and Perceived Productivity},
	author = {Meyer, André N. and Barton, Laura E. and Murphy, Gail C. and Zimmermann, Thomas and Fritz, Thomas},
	journal = {IEEE Transactions on Software Engineering},
	volume = {43},
	number = {12},
	pages = {1178--1193},
	year = {2017},
	doi = {10.1109/TSE.2017.2656886},
}

@article{mohamed_impact_2026,
	title = {The {Impact} of {LLM}-{Assistants} on {Software} {Developer} {Productivity}: {A} {Systematic} {Review} and {Mapping} {Study}},
	issn = {1049-331X, 1557-7392},
	shorttitle = {The {Impact} of {LLM}-{Assistants} on {Software} {Developer} {Productivity}},
	doi = {10.1145/3809494},
	journal = {ACM Transactions on Software Engineering and Methodology},
	author = {Mohamed, Amr and Assi, Maram and Guizani, Mariam},
	month = apr,
	year = {2026},
}

@article{mooz_dual_2006,
	title = {10.2.1 {The} {Dual} {Vee} - {Illuminating} the {Management} of {Complexity}},
	volume = {16},
	url = {https://onlinelibrary.wiley.com/doi/abs/10.1002/j.2334-5837.2006.tb02819.x},
	doi = {10.1002/j.2334-5837.2006.tb02819.x},
	number = {1},
	journal = {INCOSE International Symposium},
	author = {Mooz, Hal and Forsberg, Kevin},
	year = {2006},
	pages = {1368--1381},
}

@article{moradidakhel_github_2023,
	title = {{GitHub} {Copilot} {AI} pair programmer: {Asset} or {Liability}?},
	volume = {203},
	shorttitle = {{GitHub} {Copilot} {AI} pair programmer},
	url = {https://www.sciencedirect.com/science/article/pii/S0164121223001292},
	doi = {10.1016/j.jss.2023.111734},
	journal = {Journal of Systems and Software},
	author = {Moradi Dakhel, Arghavan and Majdinasab, Vahid and Nikanjam, Amin and Khomh, Foutse and Desmarais, Michel C. and Jiang, Zhen Ming (Jack)},
	month = sep,
	year = {2023},
	pages = {111734},
}

@inproceedings{mozannar_reading_2024,
	address = {New York, NY, USA},
	series = {{CHI} '24},
	title = {Reading {Between} the {Lines}: {Modeling} {User} {Behavior} and {Costs} in {AI}-{Assisted} {Programming}},
	shorttitle = {Reading {Between} the {Lines}},
	url = {https://doi.org/10.1145/3613904.3641936},
	doi = {10.1145/3613904.3641936},
	booktitle = {Proceedings of the 2024 {CHI} {Conference} on {Human} {Factors} in {Computing} {Systems}},
	publisher = {Association for Computing Machinery},
	author = {Mozannar, Hussein and Bansal, Gagan and Fourney, Adam and Horvitz, Eric},
	month = may,
	year = {2024},
	pages = {1--16},
}

@article{murphyhill_what_2021,
	title = {What {Predicts} {Software} {Developers}' {Productivity}?},
	volume = {47},
	url = {https://ieeexplore.ieee.org/document/8643844},
	doi = {10.1109/TSE.2019.2900308},
	number = {3},
	journal = {IEEE Transactions on Software Engineering},
	author = {Murphy-Hill, Emerson and Jaspan, Ciera and Sadowski, Caitlin and Shepherd, David and Phillips, Michael and Winter, Collin and Knight, Andrea and Smith, Edward and Jorde, Matthew},
	month = mar,
	year = {2021},
	pages = {582--594},
}

@inproceedings{nguyen_empirical_2022,
	address = {Pittsburgh, PA, USA},
	title = {An empirical evaluation of {GitHub} copilot's code suggestions},
	url = {https://dl.acm.org/doi/10.1145/3524842.3528470},
	doi = {10.1145/3524842.3528470},
	booktitle = {Proceedings of the 19th {International} {Conference} on {Mining} {Software} {Repositories}},
	publisher = {ACM},
	author = {Nguyen, Nhan and Nadi, Sarah},
	month = may,
	year = {2022},
	pages = {1--5},
}

@article{noda_dev_2023,
	title = {{DevEx}: {What} {Actually} {Drives} {Productivity}: {The} developer-centric approach to measuring and improving productivity},
	volume = {21},
	shorttitle = {{DevEx}},
	url = {https://dl.acm.org/doi/10.1145/3595878},
	doi = {10.1145/3595878},
	number = {2},
	journal = {Queue},
	author = {Noda, Abi and Storey, Margaret-Anne and Forsgren, Nicole and Greiler, Michaela},
	month = may,
	year = {2023},
	pages = {Pages 20:35--Pages 20:53},
}

@misc{pandey_transforming_2024b,
	title = {Transforming {Software} {Development}: {Evaluating} the {Efficiency} and {Challenges} of {GitHub} {Copilot} in {Real}-{World} {Projects}},
	author = {Pandey, Ruchika and Singh, Prabhat and Wei, Raymond and Shankar, Shaila},
	year = {2024},
	eprint = {2406.17910},
	archivePrefix = {arXiv},
	primaryClass = {cs.SE},
	url = {https://arxiv.org/abs/2406.17910},
}

@misc{peng_impact_2023,
	title = {The {Impact} of {AI} on {Developer} {Productivity}: {Evidence} from {GitHub} {Copilot}},
	author = {Peng, Sida and Kalliamvakou, Eirini and Cihon, Peter and Demirer, Mert},
	year = {2023},
	eprint = {2302.06590},
	archivePrefix = {arXiv},
	primaryClass = {cs.SE},
	url = {https://arxiv.org/abs/2302.06590},
}

@inproceedings{perry_users_2023,
	address = {New York, NY, USA},
	series = {{CCS} '23},
	title = {Do {Users} {Write} {More} {Insecure} {Code} with {AI} {Assistants}?},
	url = {https://doi.org/10.1145/3576915.3623157},
	doi = {10.1145/3576915.3623157},
	booktitle = {Proceedings of the 2023 {ACM} {SIGSAC} {Conference} on {Computer} and {Communications} {Security}},
	publisher = {Association for Computing Machinery},
	author = {Perry, Neil and Srivastava, Megha and Kumar, Deepak and Boneh, Dan},
	month = nov,
	year = {2023},
	pages = {2785--2799},
}

@inproceedings{pinto_developer_2024,
	address = {New York, NY, USA},
	series = {{CAIN} '24},
	title = {Developer {Experiences} with a {Contextualized} {AI} {Coding} {Assistant}: {Usability}, {Expectations}, and {Outcomes}},
	shorttitle = {Developer {Experiences} with a {Contextualized} {AI} {Coding} {Assistant}},
	url = {https://dl.acm.org/doi/10.1145/3644815.3644949},
	doi = {10.1145/3644815.3644949},
	booktitle = {Proceedings of the {IEEE}/{ACM} 3rd {International} {Conference} on {AI} {Engineering} - {Software} {Engineering} for {AI}},
	publisher = {Association for Computing Machinery},
	author = {Pinto, Gustavo and De Souza, Cleidson and Rocha, Thayssa and Steinmacher, Igor and Souza, Alberto and Monteiro, Edward},
	month = jun,
	year = {2024},
	pages = {81--91},
}

@inproceedings{prather_widening_2024,
	title = {The {Widening} {Gap}: {The} {Benefits} and {Harms} of {Generative} {AI} for {Novice} {Programmers}},
	volume = {1},
	url = {https://doi.org/10.1145/3632620.3671116},
	doi = {10.1145/3632620.3671116},
	booktitle = {Proceedings of the 2024 {ACM} {Conference} on {International} {Computing} {Education} {Research}},
	publisher = {ACM},
    address = {New York, NY, USA},
	author = {Prather, James and Reeves, Brent N. and Leinonen, Juho and MacNeil, Stephen and Randrianasolo, Arisoa S. and Becker, Brett A. and Kimmel, Bailey and Wright, Jared and Briggs, Ben},
	month = aug,
	year = {2024},
	pages = {469--486},
}

@misc{rasnayaka_empirical_2024,
	title = {An {Empirical} {Study} on {Usage} and {Perceptions} of {LLMs} in a {Software} {Engineering} {Project}},
	url = {http://arxiv.org/abs/2401.16186},
	doi = {10.48550/arXiv.2401.16186},
	publisher = {arXiv},
	author = {Rasnayaka, Sanka and Wang, Guanlin and Shariffdeen, Ridwan and Iyer, Ganesh Neelakanta},
	month = jan,
	year = {2024},
}

@article{razzaq_systematic_2024,
	title = {A {Systematic} {Literature} {Review} on the {Influence} of {Enhanced} {Developer} {Experience} on {Developers}' {Productivity}: {Factors}, {Practices}, and {Recommendations}},
	volume = {57},
	shorttitle = {A {Systematic} {Literature} {Review} on the {Influence} of {Enhanced} {Developer} {Experience} on {Developers}' {Productivity}},
	url = {https://doi.org/10.1145/3687299},
	doi = {10.1145/3687299},
	number = {1},
	journal = {ACM Comput. Surv.},
	author = {Razzaq, Abdul and Buckley, Jim and Lai, Qin and Yu, Tingting and Botterweck, Goetz},
	month = oct,
	year = {2024},
	pages = {13:1--13:46},
}

@article{sergeyuk_using_2025,
	title = {Using {AI}-based coding assistants in practice: {State} of affairs, perceptions, and ways forward},
	volume = {178},
	shorttitle = {Using {AI}-based coding assistants in practice},
	url = {https://www.sciencedirect.com/science/article/pii/S0950584924002155},
	doi = {10.1016/j.infsof.2024.107610},
	journal = {Information and Software Technology},
	author = {Sergeyuk, Agnia and Golubev, Yaroslav and Bryksin, Timofey and Ahmed, Iftekhar},
	month = feb,
	year = {2025},
	pages = {107610},
}

@misc{stackoverflow_ai_2024,
	title = {{AI} | 2024 {Stack Overflow Developer Survey}},
	url = {https://survey.stackoverflow.co/2024/ai},
	author = {{Stack Overflow}},
	year = {2024},
}

@misc{stackoverflow_ai_2025,
	title = {2025 {Stack Overflow Developer Survey}},
	url = {https://survey.stackoverflow.co/2025/},
	author = {{Stack Overflow}},
	year = {2025},
}

@article{storey_towards_2021,
	title = {Towards a {Theory} of {Software} {Developer} {Job} {Satisfaction} and {Perceived} {Productivity}},
	volume = {47},
	url = {https://ieeexplore.ieee.org/document/8851296},
	doi = {10.1109/TSE.2019.2944354},
	number = {10},
	journal = {IEEE Transactions on Software Engineering},
	author = {Storey, Margaret-Anne and Zimmermann, Thomas and Bird, Christian and Czerwonka, Jacek and Murphy, Brendan and Kalliamvakou, Eirini},
	month = oct,
	year = {2021},
	pages = {2125--2142},
}

@misc{vaillant_developers_2024,
	title = {Developers' {Perceptions} on the {Impact} of {ChatGPT} in {Software} {Development}: {A} {Survey}},
	shorttitle = {Developers' {Perceptions} on the {Impact} of {ChatGPT} in {Software} {Development}},
	url = {http://arxiv.org/abs/2405.12195},
	publisher = {arXiv},
	author = {Vaillant, Thiago S. and de Almeida, Felipe Deveza and Neto, Paulo Anselmo M. S. and Gao, Cuiyun and Bosch, Jan and de Almeida, Eduardo Santana},
	month = may,
	year = {2024},
}

@inproceedings{vaithilingam_expectation_2022,
	address = {New Orleans, LA, USA},
	title = {Expectation vs. {Experience}: {Evaluating} the {Usability} of {Code} {Generation} {Tools} {Powered} by {Large} {Language} {Models}},
	shorttitle = {Expectation vs. {Experience}},
	url = {https://dl.acm.org/doi/10.1145/3491101.3519665},
	doi = {10.1145/3491101.3519665},
	booktitle = {{CHI} {Conference} on {Human} {Factors} in {Computing} {Systems} {Extended} {Abstracts}},
	publisher = {ACM},
	author = {Vaithilingam, Priyan and Zhang, Tianyi and Glassman, Elena L.},
	month = apr,
	year = {2022},
	pages = {1--7},
}

@article{weber_significant_2024,
	title = {Significant {Productivity} {Gains} through {Programming} with {Large} {Language} {Models}},
	volume = {8},
	url = {https://doi.org/10.1145/3661145},
	doi = {10.1145/3661145},
	number = {EICS},
	journal = {Proc. ACM Hum.-Comput. Interact.},
	author = {Weber, Thomas and Brandmaier, Maximilian and Schmidt, Albrecht and Mayer, Sven},
	month = jun,
	year = {2024},
	pages = {256:1--256:29},
}

@misc{weisz_examining_2024,
	title = {Examining the Use and Impact of an {AI} Code Assistant on Developer Productivity and Experience in the Enterprise},
	url = {http://arxiv.org/abs/2412.06603},
	doi = {10.48550/arXiv.2412.06603},
	publisher = {arXiv},
	author = {Weisz, Justin D. and Kumar, Shraddha and Muller, Michael and Browne, Karen-Ellen and Goldberg, Arielle and Heintze, Ellice and Bajpai, Shagun},
	month = dec,
	year = {2024},
}

@misc{xu_ai_2026,
	title = {{AI}-{Assisted} {Programming} {Decreases} the {Productivity} of {Experienced} {Developers} by {Increasing} the {Technical} {Debt} and {Maintenance} {Burden}},
	url = {http://arxiv.org/abs/2510.10165},
	doi = {10.48550/arXiv.2510.10165},
	urldate = {2026-08-03},
	publisher = {arXiv},
	author = {Xu, Feiyang and Medappa, Poonacha K. and Tunc, Murat M. and Vroegindeweij, Martijn and Fransoo, Jan C.},
	month = jan,
	year = {2026},
}

@misc{yetistiren_evaluating_2023,
	title = {Evaluating the {Code} {Quality} of {AI}-{Assisted} {Code} {Generation} {Tools}: {An} {Empirical} {Study} on {GitHub} {Copilot}, {Amazon} {CodeWhisperer}, and {ChatGPT}},
	author = {Yetiştiren, Burak and Özsoy, Işık and Ayerdem, Miray and Tüzün, Eray},
	year = {2023},
	eprint = {2304.10778},
	archivePrefix = {arXiv},
	primaryClass = {cs.SE},
	url = {https://arxiv.org/abs/2304.10778},
}

@article{zhang_neurophysiological_2025,
	title = {The {Neurophysiological} {Paradox} of {AI}-{Induced} {Frustration}: {A} {Multimodal} {Study} of {Heart} {Rate} {Variability}, {Affective} {Responses}, and {Creative} {Output}},
	volume = {15},
	shorttitle = {The {Neurophysiological} {Paradox} of {AI}-{Induced} {Frustration}},
	url = {https://www.mdpi.com/2076-3425/15/6/565},
	doi = {10.3390/brainsci15060565},
	number = {6},
	journal = {Brain Sciences},
	author = {Zhang, Han and Wang, Shiyi and Li, Zijian},
	month = may,
	year = {2025},
	pages = {565},
}

@article{barke_grounded_2023,
	title = {Grounded {Copilot}: {How} {Programmers} {Interact} with {Code}-{Generating} {Models}},
	volume = {7},
	issn = {2475-1421},
	shorttitle = {Grounded {Copilot}},
	url = {https://dl.acm.org/doi/10.1145/3586030},
	doi = {10.1145/3586030},
	language = {en},
	number = {OOPSLA1},
	urldate = {2026-08-01},
	journal = {Proceedings of the ACM on Programming Languages},
	author = {Barke, Shraddha and James, Michael B. and Polikarpova, Nadia},
	month = apr,
	year = {2023},
	pages = {85--111},
}

@article{banh_copiloting_2025,
	title = {Copiloting the future: {How} generative {AI} transforms {Software} {Engineering}},
	volume = {183},
	issn = {09505849},
	shorttitle = {Copiloting the future},
	url = {https://linkinghub.elsevier.com/retrieve/pii/S0950584925000904},
	doi = {10.1016/j.infsof.2025.107751},
	language = {en},
	urldate = {2026-06-29},
	journal = {Information and Software Technology},
	author = {Banh, Leonardo and Holldack, Florian and Strobel, Gero},
	month = jul,
	year = {2025},
	pages = {107751},
}

@article{bainbridge_ironies_1983,
	title = {Ironies of automation},
	volume = {19},
	copyright = {https://www.elsevier.com/tdm/userlicense/1.0/},
	issn = {00051098},
	url = {https://linkinghub.elsevier.com/retrieve/pii/0005109883900468},
	doi = {10.1016/0005-1098(83)90046-8},
	language = {en},
	number = {6},
	urldate = {2026-07-04},
	journal = {Automatica},
	author = {Bainbridge, Lisanne},
	month = nov,
	year = {1983},
	pages = {775--779},
}

@article{simkute_ironies_2025,
	title = {Ironies of {Generative} {AI}: {Understanding} and {Mitigating} {Productivity} {Loss} in {Human}-{AI} {Interaction}},
	volume = {41},
	issn = {1044-7318, 1532-7590},
	shorttitle = {Ironies of {Generative} {AI}},
	url = {https://www.tandfonline.com/doi/full/10.1080/10447318.2024.2405782},
	doi = {10.1080/10447318.2024.2405782},
	language = {en},
	number = {5},
	urldate = {2026-08-01},
	journal = {International Journal of Human-Computer Interaction},
	author = {Simkute, Auste and Tankelevitch, Lev and Kewenig, Viktor and Scott, Ava Elizabeth and Sellen, Abigail and Rintel, Sean},
	year = {2025},
	pages = {2898--2919},
}

@techreport{sheridan_human_1978,
	address = {Fort Belvoir, VA, USA},
	title = {Human and {Computer} {Control} of {Undersea} {Teleoperators}:},
	shorttitle = {Human and {Computer} {Control} of {Undersea} {Teleoperators}},
	url = {https://apps.dtic.mil/sti/citations/tr/ADA057655},
	doi = {10.21236/ADA057655},
	urldate = {2026-08-01},
	institution = {Defense Technical Information Center},
	author = {Sheridan, Thomas B. and Verplank, William L.},
	month = jul,
	year = {1978},
}

@article{parasuraman_model_2000,
	title = {A model for types and levels of human interaction with automation},
	volume = {30},
	copyright = {https://ieeexplore.ieee.org/Xplorehelp/downloads/license-information/IEEE.html},
	issn = {10834427},
	url = {http://ieeexplore.ieee.org/document/844354/},
	doi = {10.1109/3468.844354},
	number = {3},
	urldate = {2026-08-01},
	journal = {IEEE Transactions on Systems, Man, and Cybernetics - Part A: Systems and Humans},
	author = {Parasuraman, R. and Sheridan, T.B. and Wickens, C.D.},
	month = may,
	year = {2000},
	pages = {286--297},
}

@article{endsley_out_1995,
	title = {The {Out}-of-the-{Loop} {Performance} {Problem} and {Level} of {Control} in {Automation}},
	volume = {37},
	copyright = {https://journals.sagepub.com/page/policies/text-and-data-mining-license},
	issn = {0018-7208, 1547-8181},
	url = {https://journals.sagepub.com/doi/10.1518/001872095779064555},
	doi = {10.1518/001872095779064555},
	language = {en},
	number = {2},
	urldate = {2026-08-01},
	journal = {Human Factors: The Journal of the Human Factors and Ergonomics Society},
	author = {Endsley, Mica R. and Kiris, Esin O.},
	month = jun,
	year = {1995},
	pages = {381--394},
}

@misc{dhanorkar_human_2026,
	title = {Human oversight of agentic systems in practice: {Examining} the oversight work, challenges, and heuristics of developers using software agents},
	copyright = {Creative Commons Attribution 4.0 International},
	shorttitle = {Human oversight of agentic systems in practice},
	url = {https://arxiv.org/abs/2606.05391},
	doi = {10.48550/ARXIV.2606.05391},
	urldate = {2026-08-01},
	publisher = {arXiv},
	author = {Dhanorkar, Shipi and Passi, Samir and Vorvoreanu, Mihaela},
	year = {2026},
}

@inproceedings{stray_developer_2026,
	title = {Developer {Productivity} {With} and {Without} {GitHub} {Copilot}: {A} {Longitudinal} {Mixed}-{Methods} {Case} {Study}},
	shorttitle = {Developer {Productivity} {With} and {Without} {GitHub} {Copilot}},
	url = {https://hdl.handle.net/10125/112286},
	doi = {10.24251/HICSS.2026.880},
	urldate = {2026-08-02},
	author = {Stray, Viktoria and Brandtzæg, Elias Goldmann and Wivestad, Viggo and Barbala, Astri and Moe, Nils Brede},
	year = {2026},
    booktitle={Proceedings of the 59th Hawaii International Conference on System Sciences (HICSS)},
    pages={7413--7422},
}

@misc{sergeyuk_evolving_2026,
	title = {Evolving with {AI}: {A} {Longitudinal} {Analysis} of {Developer} {Logs}},
	shorttitle = {Evolving with {AI}},
	url = {http://arxiv.org/abs/2601.10258},
	doi = {10.1145/3744916.3787811},
	urldate = {2026-08-02},
	author = {Sergeyuk, Agnia and Huang, Eric and Karaeva, Dariia and Serova, Anastasiia and Golubev, Yaroslav and Ahmed, Iftekhar},
	month = mar,
	year = {2026},
}

@inproceedings{ziegler_productivity_2022b,
	address = {New York, NY, USA},
	title = {Productivity assessment of neural code completion},
	isbn = {978-1-4503-9273-0},
	url = {https://dl.acm.org/doi/10.1145/3520312.3534864},
	doi = {10.1145/3520312.3534864},
	language = {en},
	booktitle = {Proceedings of the 6th {ACM} {SIGPLAN} {International} {Symposium} on {Machine} {Programming}},
	publisher = {ACM},
	author = {Ziegler, Albert and Kalliamvakou, Eirini and Li, X. Alice and Rice, Andrew and Rifkin, Devon and Simister, Shawn and Sittampalam, Ganesh and Aftandilian, Edward},
	month = jun,
	year = {2022},
	pages = {21--29},
}

@misc{cognitive_inclusivity_copilot,
      title={Cognitive Inclusivity in AI Coding Assistants: Evidence from GitHub Copilot}, 
      author={Grace Kerr and Aarushi Parasrampuria and Faith Culas and Kelly Blincoe},
      year={2026},
      eprint={2610.41192},
      archivePrefix={arXiv},
      primaryClass={cs.SE},
      url={https://arxiv.org/abs/2610.41192}, 
}



\vfill

\end{document}